\documentclass[pre,twocolumn,showpacs,preprintnumbers,amsmath,amssymb]{revtex4}

\usepackage{mathrsfs}
\usepackage{graphicx}
\usepackage{dcolumn}
\usepackage{bm}
\usepackage{array}

\begin{document}
\title{Nonequilibrium Thermodynamics of Feedback Control}
\author{Takahiro Sagawa$^{1,2}$}
\author{Masahito Ueda$^{3,4}$}
\affiliation{
$^1$ The Hakubi Center, Kyoto University, Yoshida-ushinomiya cho, Sakyo-ku, Kyoto 606-8302, Japan \\
$^2$ Yukawa Institute for Theoretical Physics, Kyoto University, Kitashirakawa-oiwake cho, Sakyo-ku, Kyoto 606-8502, Japan \\
$^3$Department of Physics, University of Tokyo, 7-3-1, Hongo, Bunkyo-ku, Tokyo, 113-8654, Japan \\
$^4$ERATO Macroscopic Quantum Control Project, JST, 2-11-16 Yayoi, Bunkyo-ku, Tokyo 113-8656, Japan
}
\date{\today}

\begin{abstract}
We establish a general theory of feedback control on classical stochastic thermodynamic systems,  and generalize nonequilibrium equalities such as the fluctuation theorem and the Jarzynski equality in the presence of feedback control with multiple measurements.  Our results are generalizations of the previous relevant works to the situations with general measurements and multi-heat baths.  The obtained equalities involve  additional terms that characterize the information obtained by measurements or the efficacy of feedback control.  A generalized  Szilard engine and a feedback-controlled ratchet are shown to satisfy the derived equalities.
\end{abstract}


\maketitle

\section{Introduction}

Since the mid-twentieth century, feedback control has played crucial roles in science and engineering~\cite{Doyle,Astrom}. Here, ``feedback'' means that a control protocol depends on measurement outcomes obtained from the controlled system.  Recently, feedback control has become increasingly important in terms of nonequilibrium physics, due to at least the following two reasons.  

First of all, stochastic aspects of  thermodynamics~\cite{Sekimoto,Bustamante,Seifert1} have become  important due to recent theoretical and experimental developments.  Theoretically, a number of nonequilibrium equalities such as the fluctuation theorem and the Jarzynski equality~\cite{Evans,Gallavotti,Jarzynski1,Crooks1,Crooks2,Spohn,Maes1,Maes2,Jarzynski2,Kurchan,Tasaki2,Hatano,Evans2,Cohen,Jarzynski3,Jarzynski4,Harada,Seifert2,EspositoQ,Gaspard,Saito,Ohkuma,Kawai,Marin,Nakagawa1,Nakagawa2,Utsumi,HanggiQ,Ren,Hasegawa,Esposito,HanggiQ2,Vaikuntanathan} have recently been found.  On the other hand,  experimental techniques have been developed to manipulate and observe small thermodynamic systems such as macromolecules and colloidal particles,  and several nonequilibrium equalities have been experimentally verified~\cite{Wang,Liphardt,Trepagnier,Carberry,Collin,Douarche,Andrieux1,Toyabe1,Toyabe2,Hayashi,Nakamura}.  Moreover,   artificial~\cite{Serreli,Rahav,Kay,H} and biological~\cite{Schliwa} molecular machines have been investigated.
In these contexts, feedback control is useful to realize intended dynamical properties of small thermodynamic systems, and it  has become a topic of active research~\cite{Cao1,Kim,Lopez,Cao2,Cao3,Bonaldi,Suzuki1,Sagawa-Ueda3,Suzuki2,Brandes,Ponmurugan,Horowitz,Morikuni,Seifert3,Horowitz2,Esposito2,Vaikuntanathan2,Sagawa,Lahiri,Horowitz3,Pekola,Toyabe3}.  

Secondly, feedback control sheds light on the foundations of thermodynamics and statistical mechanics concerning ``Maxwell's demon''~\cite{Maxwell,Demon,Szilard,Brillouin,Bennett,Landauer}.  In fact, Maxwell's demon performs measurement and feedback control on thermodynamic systems.  Recently, Maxwell's demon has attracted renewed interest~\cite{Lloyd1,Lloyd2,Lloyd3,Touchette,Zurek,Scully,Kieu,Allahverdyan,Quan,Nielsen,Sagawa-Ueda1,Jacobs,Sagawa-Ueda2,Maruyama,SWKim} from the standpoints of modern information theory and statistical mechanics.

A quintessential model of Maxwell's demon is a single-particle heat engine proposed by L. Szilard in 1929~\cite{Szilard}.  During the thermodynamic cycle of the Szilard engine, the demon obtains $1$ bit ($= \ln 2$ nat) of information by a measurement, performs feedback control, and extracts $k_{\rm B}T \ln 2$ of positive work from a single heat bath.  After numerous arguments on the consistency between the demon and the second law of thermodynamics, it is now understood that the work needed for the demon (or equivalently the feedback controller) during the measurement and information erasure compensates for the work that can be extracted by the demon~\cite{Sagawa-Ueda2}.  Therefore, we cannot extract a net positive work from the total system of the engine and the demon in an isothermal cycle, and  therefore the presence of the demon does not contradict the second law of thermodynamics.  
Nevertheless, $k_{\rm B}T \ln 2$ of work extracted by the demon can be still useful.  By using feedback control, we can increase the system's free energy without injecting any energy (work) to it.   We stress that, without feedback control, we need the direct energy input into the system in order to increase its free energy due to the second law of thermodynamics.
Feedback control may be regarded as a powerful tool to control thermodynamic systems.  
Since the crucial quantity is the information that is obtained to be used for feedback control, we may regard the Szilard-type heat engine as ``information heat engine.''
Recently, such an information heat engine was realized experimentally by using a colloidal particle~\cite{Toyabe3}.  

In this paper, we formulate a general theory of feedback control on stochastic  thermodynamic systems.  In particular, we extend recent theoretical results on the generalizations of the fluctuation theorem and the Jarzynski equality~\cite{Sagawa-Ueda3} to the situations in which the measurement and feedback control are non-Markovian and there are multi-heat baths.  Our results serve as the fundamental building blocks of information heat engines.

This paper is organized as follows.

In Sec.~II, we briefly review the framework of stochastic thermodynamics in a general setup.  We discuss classical stochastic systems that are in general non-Markovian and in contact with multi-heat baths.  
We discuss the concept of entropy production and the detailed fluctuation theorem as our starting point. 
Because they are general properties of nonequilibrium systems, our formulations and  results in the following sections are not restricted to Langevin systems but applicable to any classical stochastic systems that satisfy the detailed fluctuation theorem.

In Sec.~III, we formulate measurements on thermodynamic systems.  We discuss multi-measurements including continuous measurements, and investigate the properties of the mutual information obtained by the measurements.  In particular, we introduce the two kinds of mutual information $I$ and $I_{\rm c}$, which will be shown to play key roles in the discussion of feedback control.

In Sec.~IV, we discuss feedback control on Markov and non-Markov processes, and investigate feedback control in terms of probability theory, where the causality of the measurement and feedback play a crucial role.  

In Sec.~V, we derive the main results of this paper. 
We generalize the nonequilibrium equalities to situations in which the system is subject to feedback control.  In particular, we derive  two types of generalizations of the  fluctuation theorem and the Jarzynski equality.  
One involves a term concerning the mutual information, and the other involves a term of feedback efficacy.  As corollaries, we derive the generalizations of the second law of thermodynamics and a fluctuation-dissipation relation.

In Sec.~VI, we illustrate our general results by two examples: a generalized Szilard engine with measurement errors and a feedback-controlled ratchet~\cite{Cao1,Lopez,Cao3}.  We discuss the former analytically and the latter numerically.

In Sec.~VII, we conclude this paper.

In Appendix A, we discuss the physical meaning of entropy production to elucidate the physical contents of our results in two typical situations.

\section{Review of Stochastic thermodynamics}

In this section, we briefly review thermodynamics of classical stochastic systems and introduce notations that will be used later.

\subsection{Dynamics}

We consider a classical stochastic system $\textbf{S}$ that is in contact with heat baths $\textbf{B}_1$, $\textbf{B}_2$, $\cdots$, $\textbf{B}_n$ at respective temperatures $T_1 = (k_{\rm B}\beta_1)^{-1}$, $T_2 = (k_{\rm B}\beta_2)^{-1}$, $\cdots$, $T_n = (k_{\rm B}\beta_n)^{-1}$.  
Let $x$ be the phase-space point of system $\textbf{S}$ and $\lambda$ be a set of external parameters such as the volume of a gas or the frequency of an optical tweezers.
We control the system from time $0$ to $\tau$ with control protocol $\lambda (t)$.  Let $x(t)$ be a trajectory of the system.  

To formulate the stochastic dynamics, we discretize the time interval $[0, \tau]$ by dividing it into $N$ small intervals with width $\Delta t := \tau / N$.  The original continuous-time dynamics is recovered by taking the limit of  $N \to \infty$ or equivalently $\Delta t \to 0$.  
Let $t = n \Delta t$ and $x_n := x(n \Delta t)$.  We refer to ``time $t$'' as ``time $t_n := n \Delta t$.''  
Then, trajectory $\{ x(t') \}_{t' \in [0, t]}$ corresponds to $X_n := (x_0, x_1, \cdots, x_n)$.

Control protocol $\lambda (t)$ can also be discretized.  Let  $\lambda_n$ be the value of  $\lambda$ between $t_n = n\Delta t$ and $t_{n+1} = (n+1)\Delta t$, where it is assumed to be constant during this time interval (see FIG.~1).  
We denote the trajectory of $\lambda$ from time $0$ to $t_n$  as $\Lambda_n := (\lambda_0, \lambda_1, \cdots, \lambda_{n-1})$. 
Let $\lambda_{\rm int}$ be the value of parameter $\lambda$ before time $0$, which is not necessarily equal to $\lambda_0$ because we can switch the value of the parameter at time $0$.  
We also denote the value of $\lambda$ after time $t_N := \tau$ as $\lambda_{\rm fin}$, which is not necessarily equal to $\lambda_N$, either (see also FIG.~1).

Let $P_n[x_n]$ be the probability distribution of $x$ at time $t_n$.  In particular, $P_0 [x_0]$ is the initial distribution of $x$.  
The initial distribution can be chosen as a stationary distribution under external parameters $\lambda_{\rm int}$,  as $P_{\rm s}[x_0 | \lambda_{\rm int}]$, which means $P_0 [x_0] = P_{\rm s}[x_0 | \lambda_{\rm int}]$.   We note that $P_{\rm s}[x_0 | \lambda_{\rm int}]$  is not necessarily a canonical distribution; it can be a nonequilibrium stationary distribution.
Due to the causality, $x_{n+1}$ is determined by $X_n$ through the transition probability $P[x_{n+1} | X_n, \lambda_n]$, which depends on the external parameters at time $t_n$ (i.e., $\lambda_n$).  We note that $P[x_{n+1} | X_n, \lambda_n]$ represents the the probability of realizing $x_{n+1}$ at time $t_{n+1}$ under the condition that the trajectory of $x$ up to time $t_n$ is given by $X_n$.  If the dynamics is Markovian,  $P[x_{n+1} | X_n, \lambda_n]$ can be replaced by $P[x_{n+1} | x_n, \lambda_n]$.

\begin{figure}[htbp]
 \begin{center}
  \includegraphics[width=75mm]{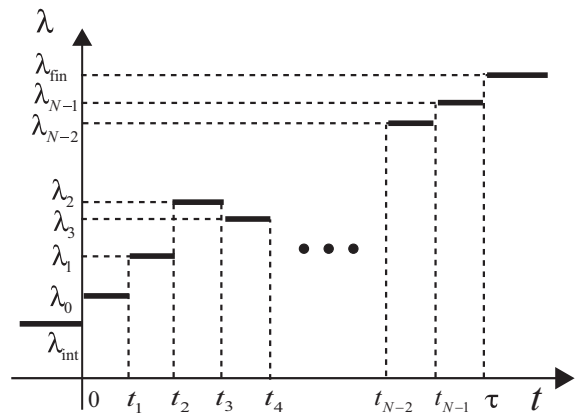}
 \end{center}
 \caption{Discretization of control protocol  $\lambda (t)$.} 
\end{figure}

The probability of trajectory $X_n$ is then given by 
\begin{equation}
P[X_n | \Lambda_n] = \prod_{k=0}^{n-1} P[x_{k+1} | X_k, \lambda_k] P_0[x_0] =: P[X_n ],
\label{prob_tr1}
\end{equation}
where we write  $P[X_n | \Lambda_n]$ just as $P[X_n ]$ for simplicity.
We note that
\begin{equation}
P[X_n | x_0, \Lambda_n] = \prod_{k=0}^{n-1} P[x_{k+1} | X_k, \lambda_k] =: P[X_n | x_0] 
\end{equation}
is the probability of trajectory $X_n$ under the condition that the initial state is $x_0$ and the control protocol is $\Lambda_n$.

Let $A$ be an arbitrary physical quantity that can depend on the trajectory $X_N$ and protocol $\Lambda_N$.  The ensemble average of this quantity is given by
\begin{equation}
\langle A \rangle = \int dX_N P[X_N | \Lambda_N] A[X_N, \Lambda_N],
\label{ensemble_average}
\end{equation}
where $dX_N := \prod_{n=0}^N dx_n$.

\subsection{Backward Control}

Before proceeding to the nonequilibrium equalities, we consider the stochastic dynamics with a backward control protocol.  The backward control protocol means the time-reversal of protocol $\Lambda_N$, which is formulated as follows.  Let $\lambda^\ast$ be the time-reversal of $\lambda$; for example, if $\lambda$ is a magnetic field, then $\lambda^\ast = - \lambda$.  The time-reversed protocol of $\lambda (t)$ is then given by $\lambda^\dagger (t) := \lambda^\ast (\tau - t)$.
The backward protocol can be discretized as  $\Lambda^\dagger_n := (\lambda^\ast_{N-1}, \lambda^\ast_{N-2}, \cdots, \lambda^\ast_{N-n-1})$.  We define $\lambda^\dagger_n := \lambda^\ast_{N-n-1}$, $\lambda^\dagger_{\rm int} := \lambda_{\rm fin}^\ast$, and $\lambda^\dagger_{\rm fin} := \lambda_{\rm int}^\ast$.

We consider the probability of realizing trajectory $x'(t)$  of the system with a backward control protocol.   We define $x'_n := x'(n \Delta t)$ and  $X'_n := (x'_0, x'_1, \cdots, x'_N)$.  We denote as $P_0^\dagger [x'_0]$ the initial distribution of the backward processes.  We stress that $P_0^\dagger [x'_0]$ is not necessarily equal to the final distribution of the forward experiments.  In fact,   we can  prepare a new state for the system to perform the backward experiments after the forward experiments.  The probability distribution of trajectory $X'_n$ with backward protocol is given by
\begin{equation}
P[X'_N | \Lambda_N^\dagger] = \prod_{k=0}^{N-1} P[x'_k | X'_k, \lambda_k^\dagger] P_0^\dagger [x'_0] =: P^\dagger[X'_N],
\label{prob_b1}
\end{equation} 
where we write $P[X'_N | \Lambda_N^\dagger]$  as $P^\dagger[X'_N]$ for simplicity.  Correspondingly,
\begin{equation}
P[X'_N | x'_0, \Lambda_N^\dagger] = \prod_{k=0}^{N-1} P[x'_k | X'_k, \lambda_k^\dagger] =: P^\dagger [X'_N | x'_0] .
\label{prob_b2}
\end{equation}

In special cases,  the backward trajectory $X'_N$ is equal to the time-reversal of the forward trajectory $X_N$.
Let $x^\ast$ be the time-reversal of phase-space point $x$.  For example, if $x = (\bm r, \bm p)$ with $\bm r$ and $\bm p$ being the position and the momentum respectively,  we have $x^\ast := (\bm r, - \bm p)$.  The time-reversal of trajectory $X_n$ is then given by $X_n^\dagger := (x_N^\ast, x_{N-1}^\ast, \cdots, x_{N-n}^\ast)$.  With notation $x_n^\dagger := x_{N-n}^\ast$, we write $X_n^\dagger = (x_0^\dagger, x_1^\dagger, \cdots, x_n^\dagger)$. By substituting $x'_n = x_n^\dagger$ to Eqs.~(\ref{prob_b1}) and (\ref{prob_b2}), we obtain the probability of realizing a backward trajectory under the backward protocol as 
\begin{equation}
P^\dagger [X_N^\dagger ] = \prod_{k=0}^{N-1} P[x_k^\dagger | X_k^\dagger, \lambda_k^\dagger] P_0^\dagger [x_0],
\end{equation} 
where the conditional probability under initial $x_0^\dagger$ is given by
\begin{equation}
P^\dagger [X_N^\dagger | x_0^\dagger] = \prod_{k=0}^{N-1} P[x_k^\dagger | X_k^\dagger,
\lambda_k^\dagger].
\end{equation}
We note that $dX_N^\dagger = dX_N$ holds, because $dx_n = dx_n^\ast$.

\subsection{Nonequilibrium Equalities}

We now discuss nonequilibrium equalities.
Let $Q_i[X_N, \lambda_N]$ be the heat that is absorbed by the system from the $i$th heat bath satisfying $Q_i[X_N, \Lambda_N] = -Q_i[X^\dagger_N, \Lambda^\dagger_N]$.  We write $Q_i[X_N, \lambda_N]$ simply as $Q_i[X_N]$ for simplicity.  It has been established that the following equality is satisfied for stochastic thermodynamic systems~\cite{Crooks1,Crooks2,Jarzynski2,Seifert2}:
\begin{equation}
\frac{P^\dagger [X^\dagger_N | x^\ast_0]}{P[X_N | x_0]} = \exp \left( \sum_i \beta_i Q_i [X_N] \right),
\label{fluctuation1}
\end{equation}
which is referred to as the detailed fluctuation theorem (or the transient fluctuation theorem). 
This is the starting point of our research.
We can rewrite Eq.~(\ref{fluctuation1})  as
\begin{equation}
\frac{P^\dagger [X^\dagger_N]}{P[X_N ]} = e^{ -\sigma [X_N] },
\label{fluctuation2}
\end{equation}
where
\begin{equation}
\sigma [X_N] := - \ln P^\dagger_0 [x^\dagger_0] + \ln P_0 [x_0] -  \sum_i \beta_i Q_i [X_N],
\label{entropy1}
\end{equation}
which is called the entropy production along trajectory $X_N$.

Various proofs of the detailed fluctuation theorem  [Eqs.~(\ref{fluctuation1}) and (\ref{fluctuation2})] for stochastic systems have been presented, for example, in  Refs.~\cite{Crooks1,Crooks2,Seifert2} for the Markovian stochastic dynamics and in Ref.~\cite{Ohkuma} for non-Markovian Langevin systems.
A proof of Eqs.~(\ref{fluctuation1}) and (\ref{fluctuation2})  has also been given in Ref.~\cite{Jarzynski2} for the situations in which the total system including heat baths is treated as a Hamiltonian system and the initial states of the heat baths in the forward and backward processes are the canonical distributions. This proof can confirm the physical validity of the detailed fluctuation theorem even for the non-Markovian dynamics with multi-heat baths, as the stochastic dynamics can be reproduced as that of a partial system of the total Hamiltonian system including the heat baths.
We also note that several equalities that are similar but not equivalent to Eqs.~(\ref{fluctuation1}) and (\ref{fluctuation2}) have been derived for different situations.  For example, the transient fluctuation theorem has been discussed for dynamical systems in Ref.~\cite{Evans2}.
The fluctuation theorem for nonequilibrium  steady states has been discussed for stochastic systems~\cite{Kurchan0,Spohn} and dynamical systems~\cite{Evans,Gallavotti}.

From the detailed fluctuation theorem  (\ref{fluctuation2}), we can show   Crooks'  fluctuation theorem as follows.  We denote as $P[\sigma ]$  the probability of finding the entropy production $\sigma$ in the forward processes, satisfying
\begin{equation}
P[\sigma ] = \int \delta (\sigma - \sigma [X_N]) P[ X_N] dX_N,
\end{equation}
where $\delta (\cdot )$ is the delta function.  On the other hand, let $P^\dagger[ \sigma] $ be the probability of obtaining $\sigma$ in the backward processes, satisfying 
\begin{equation}
P^\dagger[\sigma ] = \int \delta (\sigma - \sigma [X'_N]) P^\dagger[ X'_N] dX'_N.
\end{equation}
By using the detailed fluctuation theorem (\ref{fluctuation2}) and equality $\sigma [X_N] = - \sigma [X^\dagger_N]$, we obtain Crooks' fluctuation theorem
\begin{equation}
\frac{P^\dagger [-\sigma]}{P[\sigma]} = e^{-\sigma}.
\label{fluctuation_Crooks}
\end{equation}

The detailed fluctuation theorem (\ref{fluctuation2}) or Crooks' fluctuation theorem (\ref{fluctuation_Crooks}) leads to the integral fluctuation theorem
\begin{equation}
\langle e^{- \sigma} \rangle = 1,
\label{integral1}
\end{equation}
where the ensemble average $\langle \cdots \rangle$ is taken over all trajectories under forward protocol (see Eq.~(\ref{ensemble_average})).
From the concavity of the exponential function, we obtain
\begin{equation}
\langle \sigma \rangle \geq 0,
\label{second1}
\end{equation}
which is an expression of the second law of thermodynamics: the ensemble-averaged entropy production is non-negative.  By taking the ensemble average of the logarithm of both sides of Eq.~(\ref{fluctuation2}), we have
\begin{equation}
\langle \sigma \rangle  = \int dX_N P[X_N] \ln \frac{P[X_N]}{P^\dagger[X_N^\dagger]},
\label{relative1}
\end{equation}
which we will refer to as the Kawai-Parrondo-Broeck (KPB) equality~\cite{Kawai,Marin}.  The right-hand side of Eq.~(\ref{relative1}) is the Kullback-Leibler divergence (or the relative entropy) of $P[X_N]$ and $P^\dagger[X_N^\dagger]$, which is always positive.  Therefore, Eq.~(\ref{relative1}) reproduces inequality (\ref{second1}).

If the probability distribution of $\sigma$ is Gaussian, the cumulant expansion of Eq.~(\ref{ensemble_average}) leads to a variant of  fluctuation-dissipation relation
\begin{equation}
\langle \sigma \rangle = \frac{1}{2}(\langle \sigma^2 \rangle - \langle \sigma \rangle^2),
\label{FDT1}
\end{equation}
which indicates that  $\langle \sigma \rangle$ is determined by the fluctuation of $\sigma$.
Equality~(\ref{FDT1}) is an expression of the fluctuation-dissipation theorem of the first kind, which gives a special case of the Green-Kubo formula~\cite{Evans2}.

In the case of an isothermal process with a single heat bath, the entropy production reduces to
\begin{equation}
\sigma [X_N] = \beta (W[X_N] - \Delta F),
\end{equation}
where $W[X_N]$ is the work performed on the system during the process, and $\Delta F$ is the difference of the free energies for the initial and final Hamiltonians (see also Appendix A for details).
Under this situation, Eq.~(\ref{integral1}) leads to 
\begin{equation}
\langle e^{- \beta W} \rangle = e^{-\beta \Delta F}, 
\label{Jarzynski1}
\end{equation}
which is the Jarzynski equality~\cite{Jarzynski1}.  The second law of thermodynamics  then reduces to
\begin{equation}
\langle W \rangle \geq \Delta F.
\label{second_work}
\end{equation}

\section{Measurement}

In this section, we formulate and investigate the effect of measurements on nonequilibrium dynamics.  

\subsection{Classical Measurement and Mutual Information}

In this subsection, we review the general framework of a measurement on a probabilistic variable, which can be applied to a broad class of measurements on classical systems.  

Let $x$ be an arbitrary probability variable of a measured system whose distribution is $P [x]$.  We perform a measurement on it and obtain outcome $y$ which is also a probability variable.  The error of the measurement can be characterized by a conditional probability $P[y | x]$, which describes the probability of obtaining outcome $y$ under the condition that the true value of the measured system is $x$.   We note that $\sum_y P[y | x] = 1$ for all $x$, where we note that the sum should be replaced by the integral if $y$ is a continuous variable.  If the measurement is error-free, $P[y | x]$ is given by the delta function or the Kronecker's delta. We assume that $P[y | x]$ is independent of the probability distribution $P[x]$; in other words, the error is independent of the state preparation of the measured system.  
The joint probability of $x$ and $y$ is given by $P[x,y] = P[y|x] P[x]$, and the probability of obtaining $y$ by $P[y] = \sum_x P[x, y]$.
The probability of realizing $x$ under the condition that the measurement outcome is  $y$, denoted as $P[x |y]$, is given by the Bayes theorem:
\begin{equation}
P[x|y] = \frac{P[y|x]P[x]}{P[y]}.
\label{Bayes}
\end{equation}

We next discuss the information contents related to the measurement~\cite{Shannon,Cover-Thomas}.
The Shannon information contents of the probability variables are given by
\begin{equation}
H_x := -\sum_x P[x] \ln P[x], \ H_y := - \sum_y P[y] \ln P[y],
\end{equation}
which characterize the randomnesses of $x$ and $y$, respectively.  On the other hand, the mutual information content $\langle I \rangle$ between $x$ and $y$ is given by
\begin{equation}
\langle I \rangle := \sum_{xy} P[x,y]I[x:y],
\end{equation}
where
\begin{equation}
I[x:y] := \ln \frac{P[y|x]}{P[y]}.
\end{equation}
In this paper, we also call $I[x:y]$ the mutual information.  We note that $I[x:y] = I[y:x]$ holds due to the Bayes theorem (\ref{Bayes}).

The mutual information $\langle I \rangle$ measures the amount of information obtained by the measurement.    It is known that
\begin{equation}
0 \leq \langle I \rangle \leq H_x, \ 0 \leq \langle I \rangle \leq H_y.
\end{equation}
If the measurement is error-free, $\langle I \rangle = H_x = H_y$ holds.

\subsection{Measurements on Nonequilibrium Dynamics}

We next  formulate multiple measurements on nonequilibrium dynamics, and discuss the properties of the mutual information obtained by the measurements.

Let $y_n$ be the outcome at time $t_n := n \Delta t$.
In this section, we assume the followings:
\begin{enumerate}
\item The error of the measurement at time $t_n$ is characterizes by $P[y_n | X_n]$, where $y_n$ can depend on the trajectory of the system before $t_n$ due to the causality.     Here we assumed that the property of the measurement error at time $t_n$ does not explicitly depend on $Y_{n-1}$ or $P[X_n]$.  This assumption is also justified in many real experimental situations. 

\item The unconditional probability distribution of $X_n$, $P[X_n]$, is not affected by the back-action of the measurement.  Since the system is classical, this assumption is justified for many real systems  such as colloidal particles and macromolecules. 
\end{enumerate}

If $P[y_n | X_n] = P[y_n | x_n]$, we call the measurement Markovian, which means that the outcome is determined only by the system's state immediately before the measurement.  This condition is satisfied if the measurements can be performed in a time interval that is sufficiently shorter than the shortest time scale $\Delta t$ of the system.  We note that the Markovness of the measurement is independent of that of the dynamics.

We assume that the measurements are performed at times $t_{n_1}$, $t_{n_2}$, $\cdots$, $t_{n_M}$, where $0 \leq n_1 < n_2 < \cdots < n_M \leq N$.  If $n_1 = 0$, $n_2 = 1$, $n_3 = 2$, $\cdots$, $n_{N+1} = N$ hold, the measurement is time-continuous in the limit of $\Delta t \to 0$, because the measurements are performed  at all times. 

We write as $Y_n$ the set of measurement outcomes that are obtained up to time $t_n$, i.e. $Y_n := (y_{n_1}, y_{n_2}, \cdots, y_{[n]})$ where $[n]$ is the maximum $n_k$ satisfying $n_k \leq n$.  If the measurement is continuous, then $Y_n = (y_0, y_1, \cdots, y_n)$.

We define
\begin{equation}
P_{\rm c}[Y_n | X_n] := \prod_{k=1}^{M'} P[y_{n_k} | X_{n_k}],
\label{c}
\end{equation}
where $M'$ is the maximum integer satisfying $n_{M'} \leq n$.  
Without feedback, Eq.~(\ref{c}) defines the conditional probability of obtaining outcomes $Y_n$ under the condition of $X_n$, while, with feedback, this interpretation of Eq.~(\ref{c}) is not necessarily correct as shown in the next section.  To explicitly demonstrate this point and to distinguish $P_{\rm c}[Y_n | X_n]$ from the usual conditional probability,  we put suffix ``c''.
Then, the joint distribution of $X_n$ and $Y_n$  is given by
\begin{equation}
P[X_n, Y_n] = P_{\rm c}[Y_n | X_n] P[X_n].
\end{equation}
The probability of obtaining outcomes $Y_n$ is given by
\begin{eqnarray}
P[Y_n ] = \int dX_n P[X_n, Y_n] = \prod_{k=1}^{M'} P[y_{n_k} | Y_{n_k-1}],
\label{prob0}
\end{eqnarray}
where the two equalities are just identities known in probability theory.
We also note that
\begin{equation}
\begin{split}
P[y_n | X_n, Y_{n-1}] &:= \frac{P[Y_n | X_n]}{P[Y_{n-1} | X_n]} \\
&= P[y_n | X_n],
\end{split}
\label{prob1}
\end{equation}
which is, in fact, independent of $Y_{n-1}$.

We then discuss the mutual information obtained by multiple measurements on nonequilibrium dynamics.  Suppose that we obtain measurement outcomes $Y_{n-1}$ at time $t_{n-1}$.  If we perform another measurement at time $t_n$ and obtain outcome $y_n$, we obtain the mutual information between $y_n$ and $X_n$ under the condition that we have obtained $Y_{n-1}$:
\begin{equation}
\begin{split}
I[y_n : X_n | Y_{n-1}] &:= \ln \frac{P[y_n | X_n, Y_{n-1}]}{P[y_n | Y_{n-1}]} \\
&= \ln \frac{P[y_n | X_n]}{P[y_n | Y_{n-1}]},
\end{split}
\end{equation}
where  we used Eq.~(\ref{prob1}). 
 We note that, if the measurement is Markovian, $I[y_n : X_n | Y_{n-1}]$ reduces to $I[y_n : x_n | Y_{n-1}]$.
We denote as $I_{\rm c}$ the sum of these mutual information contents obtained by multi-measurements, that is,
\begin{equation}
\begin{split}
I_{\rm c} [ X_n: Y_n] &:= \sum_{k=1}^{M'} I [y_{n_k}: X_{n_k}] \\
&= \ln \frac{P[Y_n | X_n]}{P[Y_n]},
\end{split}
\label{mutual1}
\end{equation} 
where we used Eq.~(\ref{prob0}).  We note that the same quantity has been discussed in Ref.~\cite{Cao2}.
From Eq.~(\ref{mutual1}), we find that $I_{\rm c} [Y_n : X_n]$ equals the mutual information between trajectories $X_n$ and $Y_n$ defined as $I [Y_n : X_n ] := \ln (P[Y_n | X_n] / P[Y_n])$.  In the presence of feedback control, however, this is not true (i.e., $I_{\rm c} \neq I$), as we will see later.

\section{Feedback Control}

In this section, we formulate feedback control on nonequilibrium dynamics.

\subsection{Formulation}

Feedback control implies that  protocol $\Lambda_N$ depends on measurement outcomes $Y_N$ (see FIG.~2).   On the other hand, without feedback control, control protocols  are predetermined and independent of the measurement outcomes, as is the case for the setup of the original fluctuation theorem and Jarzynski equality. 

\begin{figure}[htbp]
 \begin{center}
  \includegraphics[width=75mm]{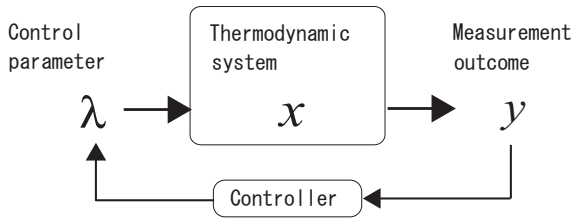}
 \end{center}
 \caption{Feedback control on nonequilibrium dynamics.  The control parameter  is denoted as $\lambda$, the point of the phase space of the system as $x$, and the outcome of measurement on the system as $y$.  Parameter $\lambda$ depends on  $y$ through the real-time feedback control. } 
\end{figure}

When the system is subject to feedback control, $\lambda_n$ can depend on measurement outcomes that are obtained until  $t_n$, while $\lambda_n$ cannot depend on  any measurement outcome that is obtained after time $t_n$ due to causality.  
We introduce notation $\lambda_n (Y_n)$, which means that the value of $\lambda$ at time $t_n$ is determined  by $Y_n$.   We write $\Lambda_n (Y_{n-1}) := (\lambda_0 (Y_0), \lambda_1 (Y_1), \cdots, \lambda_{n-1} (Y_{n-1}))$. 

If $\lambda_n$ only depends on $y_n$ as $\lambda_n (y_n)$, the feedback protocol is called Markovian.  We note that the Markovness of feedback is independent of that of the dynamics or measurements.  The Markovian feedback control is realized when the delay time of feedback is sufficiently smaller than the smallest time  scale $\Delta t$ of the dynamics.

\subsection{Overdamped Langevin System}

As a simple illustrative example, we discuss an overdamped Langevin system, whose equation of motion is given by
\begin{equation}
\eta \frac{dx (t)}{dt} = -\frac{\partial V(x, \lambda)}{\partial x} + f(\lambda) + \sqrt{2 \eta k_{\rm B}T} \xi (t),
\label{Langevin}
\end{equation}
where $\eta$ is the friction constant, $V(x, \lambda)$ is an external potential, $f(\lambda)$ is an external nonconservative force, and $\xi (t)$ is the Gaussian white noise satisfying $\langle \xi (t) \xi (t') \rangle = \delta (t-t')$.  
The detailed fluctuation theorem (\ref{fluctuation1}) is still satisfied in the presence of a nonconservative force that violates the detailed balance,  because Eq.~(\ref{fluctuation1})  can be derived from the local transition rate of the stochastic dynamics and is independent whether there is a global potential or not. 
We assume that the measurement is time-continuous and Markovian:
\begin{equation}
y_n = x_n + \frac{\Delta R_n}{\Delta t },
\label{measurement_Langevin}
\end{equation}
where $\Delta R_n$ is a white Gaussian noise with $\langle \Delta R_n \Delta R_{n'} \rangle = R \delta_{nn'}\Delta t$ ($R>0$).  The conditional probability of obtaining outcome $y_n$ is given by
\begin{equation}
P[y_n | x_n] \propto \exp \left[  - \frac{\Delta t}{2R} (y_n -x_n)^2 \right].
\end{equation}

The feedback protocol can be written as $\lambda_n (y_0, y_1, \cdots, y_n)$ in general.
The work performed on the system is then given by
\begin{equation}
\begin{split}
W_n &:= V(x_n, \lambda_{n+1}) - V (x_n, \lambda_n) \\
&= \frac{\partial V}{\partial \lambda} \Delta \lambda_n + o(\Delta t), 
\end{split}
\end{equation}
where 
\begin{equation}
\Delta \lambda_n := \lambda_{n+1} (y_0, y_1, \cdots, y_{n+1}) - \lambda_n (y_0, y_1, \cdots, y_n).
\end{equation}  
In particular, if the feedback is Markovian, $\lambda_n$ is given by $\lambda_n (y_n)$.  Then,  $\Delta \lambda_n = \lambda_{n+1} (y_{n+1}) - \lambda_n (y_n)$ can be written as
\begin{equation}
\Delta \lambda_n = \frac{\partial \lambda}{\partial t} \Delta t + \frac{\partial \lambda}{\partial y} \Delta y_n + \frac{1}{2}\frac{\partial^2 \lambda}{\partial y^2} \Delta y_n^2,
\label{lambda}
\end{equation}
where $\Delta y_n := y_{n+1} - y_n$.   The first term on the right-hand side of Eq.~(\ref{lambda}) arises from a change in  $\lambda$ by the pre-fixed protocol, while the second and third terms are induced by the feedback control.

We next consider  the Kalman filter and the optimal control.  As a special case of Eq.~(\ref{Langevin}), we consider a discretized linear Langevin equation in the It\^{o} form:
\begin{equation}
\eta (x_{n+1} - x_n) = - K x_n \Delta t + \lambda_n \Delta t + \sqrt{2 \eta k_{\rm B}T} \Delta W_n,  
\label{linear_Langevin}
\end{equation}
where $K$ is a positive constant and $\lambda_n$ is a control parameter.
If the initial distribution of $x_0$ is Gaussian, the distribution of $x_n$ remains Gaussian with Eq.~(\ref{linear_Langevin}).
In this case, the obtained mutual information by measurement (\ref{measurement_Langevin}) at time $t_n$ is given by
\begin{equation}
\langle I_{\rm c} [x_n : y_n] \rangle = \frac{1}{2} \ln \left(1 + \frac{S_n}{R}\Delta t \right) = \frac{S_n}{2R}\Delta t + o(\Delta t),
\end{equation}
where $S_n := \langle x_n^2 \rangle - \langle x_n \rangle^2$.  Therefore, the total mutual information
\begin{equation}
\langle I_{\rm c} \rangle = \lim_{N \to \infty} \sum_{n=0}^{N} \frac{S_n}{2R}\Delta t
\end{equation}
can converge, while the measurement is continuous.

We consider the Kalman filter on Eq.~(\ref{linear_Langevin}) with measurement (\ref{measurement_Langevin}).  The Kalman filter is a standard method to construct the optimal estimator of $x_n$, denoted as $\hat x_n$, in terms of the mean square error.   From measurement outcomes $Y_n$, $\hat x_n$ is obtained as the solution to the following simultaneous differential equations~\cite{Welch}:
\begin{equation}
\hat x_{n+1} - \hat x_n = - \frac{K \hat x_n + \lambda_n}{\eta}  \Delta t + \frac{A_n}{R} (y_n - \hat x_n )\Delta t, 
\end{equation}
\begin{equation}
A_{n+1} - A_n = \left(  \frac{-2K A_n + 2k_{\rm B}T} {\eta}- \frac{A_n}{R} \right)\Delta t, 
\label{R}
\end{equation}
where $A_n$ is a time-dependent real number and Eq.~(\ref{R}) is a discretized version of the Riccati equation.  By using the Kalman estimator $\hat x_n$, the optimal control protocol~\cite{Bertsekas} is given by
\begin{equation}
\lambda_n = - C_n \hat x_n,
\end{equation}
where $C_n$ is a pre-determined constant depending on the target of the optimal control.   We note that the optimal control is a non-Markovian control as $\lambda_n = \lambda_n (Y_n)$, because we use all of $Y_n = (y_0, y_1, \cdots, y_n)$ to calculate $\hat x_n$.
The generalized Jarzynski equality for this situation has been discussed in Ref.~\cite{Suzuki2}.


\subsection{Probability Distributions with Feedback}

We discuss the probability distributions with feedback control in general.
Under the condition that  we fix  control protocol $\Lambda_N (Y_N)$ with $Y_N$ being fixed, the conditional probability of realizing $X_N$ is given by
\begin{equation}
P[X_n | \Lambda_n (Y_{n-1})] = P_0 [X_0] \prod_{k=0}^{k-1} P[x_{k+1} | X_k, \lambda_k (Y_k)],
\label{prob_tr_f}
\end{equation}
which corresponds to Eq.~(\ref{prob_tr1}).
We note that, in the expression in Eq.~(\ref{prob_tr_f}), we do not omit the notation $\Lambda_N (Y_{N-1})$ because its $Y_{N-1}$-dependence is crucial.  We also write 
\begin{equation}
P[X_n | x_0, \Lambda_n (Y_{n-1})] := \prod_{k=0}^{k-1} P[x_{k+1} | X_k, \lambda_k (Y_k)].
\end{equation}
On the other hand, along the trajectory $X_n$, the conditional probability of obtaining outcome $y_n$ at time $t_n$ is written as $P[y_n | X_n]$.  We then define
\begin{equation}
P_{\rm c}[Y_n | X_n] := \prod_{k=0}^{n-1} P[y_k | X_k],
\end{equation}
which is to be compared with Eq.~(\ref{c}).

We then obtain the joint probability distribution of $X_n$ and $Y_n$ with feedback control as
\begin{equation}
\begin{split}
P[X_n, Y_n] &= \prod _{k=0}^{n-1} P[y_{k+1} | X_{k+1}] P[x_{k+1} | X_{k}, \lambda_{k} (Y_{k-1})] \\
&= P_{\rm c}[Y_n | X_n] P [X_n | \Lambda_n (Y_{n-1})]. 
\end{split}
\label{joint_prob}
\end{equation}
We can check that
\begin{equation}
\int P[X_n, Y_n] dX_n dY_n = 1,
\end{equation}
by integrating $X_n$ and $Y_n$ in Eq.~(\ref{joint_prob}) in the order of $y_n \to x_n \to y_{n-1} \to x_{n-1} \to \cdots \to y_1 \to x_1 \to y_0 \to x_0$, where the causality of measurements and feedback play crucial roles.

The marginal distributions are given by
\begin{equation}
P[X_n] = \int P[X_n, Y_n] dY_n, \ P[Y_n] = \int P[X_n, Y_n] dX_n,
\end{equation}
and the conditional distributions by
\begin{equation}
P[X_n | Y_n] = \frac{P[X_n, Y_n]}{P[Y_n]},  \ P[Y_n | X_n] = \frac{P[X_n, Y_n]}{P[X_n]}.
\end{equation}
We stress that, in the presence of feedback control,
\begin{equation}
P[Y_n | X_n] \neq P_{\rm c}[Y_n | X_n] 
\end{equation}
in general, because protocol $\Lambda_N$ depends on $Y_{N-1}$.  On the other hand, without feedback control, $P[Y_n | X_n] = P_{\rm c}[Y_n | X_n]$ holds because $P[X_n]$ is simply given by $P[X_n | \Lambda_n]$ with $\Lambda_n$ being independent of $Y_n$.

The ensemble average of a probability variable $A[X_n, Y_n]$ is given by
\begin{equation}
\langle A \rangle := \int A[X_n, Y_n] P[X_n, Y_n] dX_n dY_n,
\end{equation}
and the conditional average under the condition of $Y_n$ is given by
\begin{equation}
\langle A \rangle_{Y_n} := \int A[X_n, Y_n] P[X_n | Y_n] dX_n.
\end{equation}
Equation (\ref{prob1}) still holds in the presence of feedback control:
\begin{equation}
\begin{split}
P[y_n | X_n, Y_{n-1}] &:= \frac{P[X_n, Y_n]}{P[X_n, Y_{n-1}]} \\
&= \frac{P_{\rm c}[Y_n | X_n] P[X_n | \Lambda_n (Y_{n-1})]}{P_{\rm c} [Y_{n-1} | X_{n-1}] P[X_n | \Lambda_n (Y_{n-1})]} \\
&= P[y_n | X_n].
\end{split}
\end{equation}
We note that Eq.~(\ref{prob0}) also holds with feedback control.

We then define the mutual information in the same way as in the case without feedback control:
\begin{equation}
\begin{split}
I_{\rm c} [Y_n : X_n] &:= \sum_{k=1}^{M'} I[y_{n_k}: X_{n_k} | Y_{n_k-1}] \\
&= \ln \frac{P_{\rm c}[Y_n | X_n]}{P[Y_n]}.
\end{split}
\label{mutual_f}
\end{equation}
In the presence of feedback control,  $I_{\rm c} [Y_n : X_n]$ does not equal  the mutual information between trajectories $X_n$ and $Y_n$ defined as $I [Y_n : X_n] := \ln (P[Y_n | X_n] / P[Y_n])$, because $P_{\rm c}[Y_n | X_n] \neq P[Y_n | X_n]$.  Intuitively speaking, $I_{\rm c}$ only characterizes the correlation between $X_n$ and $Y_n$ due to the measurements, while $I$ involves the correlation due to the feedback control.  Note that   $I_{\rm c}$ is a more important quantity than $I$, because $I_{\rm c}$ has a clear information-theoretic significance: $I_{\rm c}$ is the information that we obtain by measurements.  We also note that, in the case of a single measurement and feedback, $I_{\rm c} = I$ always holds.

We also note that an identity similar to the integral fluctuation theorem holds for $I_{\rm c}$:
\begin{equation}
\langle e^{- I_{\rm c}} \rangle = 1,
\label{mutual_integral}
\end{equation}
because
\begin{equation}
\begin{split}
\langle e^{- I_{\rm c}} \rangle &= \int dX_N dY_N \frac{P[Y_N]}{P_{\rm c}[Y_N | X_N]} P[X_N, Y_N] \\
&= \int dX_N dY_N P[Y_N] P[X_N | \Lambda_N (Y_{N-1})] = 1.
\end{split}
\end{equation}

\subsection{Detailed Fluctuation Theorem for a Fixed Control Protocol}

If we fix control protocol $\Lambda_N (Y_N)$  with $Y_N$ being fixed, then the detailed fluctuation theorem (\ref{fluctuation1}) still holds:
\begin{equation}
\frac{P[X^\dagger_N | x^\ast_0, \Lambda_N (Y_{N-1})^\dagger]}{P[X_N | x_0, \Lambda_N (Y_{N-1})]} = \exp \left( \sum_i \beta_i Q_i [X_N, \Lambda_N (Y_{N-1})] \right),
\label{fluctuation_f1}
\end{equation}
where
\begin{equation}
\Lambda_N (Y_N)^\dagger := (\lambda_{N-1} (Y_{N-1})^\ast, \cdots,  \lambda_0 (Y_0)^\ast).
\label{backward_control}
\end{equation}
The left-hand side of Eq.~(\ref{fluctuation_f1}) corresponds to the following forward and backward experiments.  We first perform forward experiments many times with feedback control, and choose the subensemble in which the measurement outcomes are given by $Y_{N-1}$.  Within this subensemble, the ratio of trajectory $X_N$ is given by $P[X_N | x_0, \Lambda_N (Y_{N-1})]$  under the condition of initial $x_0$.  We next perform backward experiments with protocol $\Lambda_N (Y_{N-1})^\dagger$, where $Y_{N-1}$ was  chosen in the forward experiments.  We stress that we do not perform any feedback in the backward experiments:  $\Lambda_N (Y_{N-1})^\dagger$ is just the time-reversal of $\Lambda_N (Y_{N-1})$.  We then obtain $P[X^\dagger_N | x^\ast_0, \Lambda_N (Y_{N-1})^\dagger]$ as  the ratio of  trajectory $X_N^\dagger$, under the condition of initial $x_0^\dagger$ in the backward experiments.  The original detailed fluctuation theorem (\ref{fluctuation1}) can straightforwardly be applied to this subensemble corresponding to $Y_{N-1}$ because we have a unique control protocol in the subensemble, and therefore we obtain Eq.~(\ref{fluctuation_f1}). 

Let the initial distribution of the backward experiments be $P_0^\dagger [x_0^\dagger | Y_N]$ which in general depends on the measurement outcomes in the forward experiments.  A natural choice of $P_0^\dagger [x_0^\dagger | Y_N]$ is a stationary state $P_{\rm s} [x_0^\dagger | \lambda (Y_N)^\ast]$.  Then we have
\begin{equation}
\frac{P[X^\dagger_N |  \Lambda_N (Y_N)^\dagger]}{P[X_N | \Lambda_N (Y_N)]} = \exp \left( - \sigma [X_N, \Lambda_N (Y_N)] \right),
\end{equation}
where
\begin{equation}
\begin{split}
\sigma [X_N, \Lambda_N (Y_N)] := &- \ln P^\dagger_0 [x^\dagger_0 | Y_N] + \ln P_0 [x_0] \\
&-  \sum_i \beta_i Q_i [X_N, \Lambda_N (Y_{N-1}) ].
\end{split}
\label{entropy_f1}
\end{equation}
If there is a single heat bath and the initial distributions of the forward and backward  experiments are given by the canonical distributions, then the entropy production reduces to
\begin{equation}
\sigma [X_N, \Lambda_N (Y_N)] =  \beta (W[X_N, \Lambda_N (Y_N)] - \Delta F [Y_N]),
\end{equation}
where the free-energy difference can depend on the measurement outcomes as $\Delta F [Y_N] := F(\lambda_{\rm fin} (Y_N)) - F(\lambda_{\rm int})$.

We list our notations in Table I.

\begin{table}[htbp]
\begin{tabular}{|c|m{6.0cm}|}
\hline
$x$ & Phase-space point of the system. \\
\hline
$x^\ast$ & Time-reversal of $x$. \\
\hline
$t_n := n \Delta t$ & Discretization of time. \\
\hline
$x_n$ & Phase-space point at time $t_n$. \\
\hline
$X_n$ & Trajectory of the phase-space point from time $0$ to $t_n$, i.e. $X_n := (x_0, x_1, \cdots, x_n)$. \\
\hline
$X_n^\dagger$ & Time-reversal of $X_n$, i.e. $X_n^\dagger := (x_0^\dagger, x_1^\dagger, \cdots, x_n^\dagger)$ with $x_n^\dagger := x_{N-n}^\ast$.\\
\hline
$\lambda_n$ & Controllable external parameters at time $t_n$. \\
\hline
$\Lambda_n$ & Control protocol from time $0$ to $t_n$. \\
\hline
$\Lambda_n^\dagger$ & Backward control protocol of $\Lambda_n$, i.e. $\Lambda_n^\dagger := (\lambda_0^\dagger, \lambda_1^\dagger, \cdots, \lambda_n^\dagger)$ with $\lambda^\dagger_n := \lambda^\ast_{N-n-1}$. \\
\hline
$\sigma [X_n, \Lambda_n]$ & Entropy production. \\
\hline
$y_n$ & Measurement outcome at time $t_n$. \\
\hline
$Y_n$ & Measurement outcomes from time $0$ to $t_n$. \\
\hline
$Y_n^\dagger$ & Time-reversal of $Y_n$, i.e.  $(y_0^\dagger, y_1^\dagger, \cdots, y_n^\dagger)$ with $y_n^\dagger := y_{N-n}^\ast$.\\
\hline
$\Lambda_n (Y_{n-1})$ & Protocol of feedback control with outcomes $Y_{n-1}$. \\
\hline
$P[X_n | \Lambda_n (Y_n)]$ & Probability of trajectory $X_n$ under the condition that the protocol is given by  $ \Lambda_n (Y_n)$ with $Y_n$ being fixed outcomes. \\
\hline
$\Lambda_n (Y_{n-1})^\dagger$ & Time-reversal of $\Lambda_n (Y_{n-1})$:  $\Lambda_n (Y_{n-1})^\dagger := (\lambda_{N-1} (Y_{N-1})^\ast, \cdots,  \lambda_{N-n-1} (Y_{N-n-1})^\ast).$\\
\hline
$P[y_n| X_n]$ &  Probability density of obtaining $y_n$ under the condition of $X_n$, which characterizes the measurement error.  \\
\hline
$P_{\rm c} [Y_n| X_n]$ & $:= \prod_k P[y_{n_k}| X_{n_k}]$.\\
\hline
$P[X_n, Y_n]$ & Joint distribution of $X_n$ and $Y_n$ which is given by  $P[X_n, Y_n] = P_{\rm c} [Y_n| X_n]P[X_n | \Lambda_n (Y_{n-1})]$. \\
\hline
$I[y_n : X_n | Y_{n-1}]$ & Conditional mutual information obtained at time $t_n$ under the condition that we have outcomes $Y_{n-1}$. \\
\hline
$I_{\rm c} [X_n:Y_n] $& Sum of the conditional mutual information: $I_{\rm c} [X_n:Y_n] := \prod_k I[y_{n_k} : X_{n_k} | Y_{n_k-1}]$. \\
\hline
\end{tabular}
\caption{Symbols and their meanings.}
\end{table}

\section{Nonequilibrium Equalities with Feedback Control}

We now discuss the main results of this paper.  We derive the two types of the generalized nonequilibrium equalities with feedback control in Sec.~V A and V B, respectively. The former generalization involves the mutual information, while the latter  involves the efficacy of feedback control. 

\subsection{Generalized Fluctuation Theorem with Mutual Information}

To derive a generalized detailed fluctuation theorem, we first formulate the relevant backward probabilities.   We consider the following type of ``backward probability distribution'':
\begin{equation}
P^\dagger [X_N^\dagger, Y_N] := P[X^\dagger_N |  \Lambda_N (Y_{N-1})^\dagger] P[Y_N],
\label{prob_b}
\end{equation}
which satisfies
\begin{equation}
\int P^\dagger [X_N^\dagger, Y_N   ] dX_N^\dagger dY_N^\dagger = 1.
\end{equation}
Definition (\ref{prob_b}) has a clear operational meaning.  Suppose that we perform a forward experiment with feedback and obtain outcome $Y_N$.  We then perform a backward experiment with protocol $\Lambda_N (Y_{N-1})^\dagger$.  We repeat this set of the forward and backward experiments many times, and calculate the fractions of $(X_N, Y_N)$ and $(X_N^\dagger, Y_N)$, which respectively  give $P [X_N, Y_N]$ and $P^\dagger [X_N^\dagger, Y_N]$.

Noting  Eq.~(\ref{joint_prob}) and the definition of the mutual information (\ref{mutual_f}),  we obtain a generalized detailed fluctuation theorem with feedback control:
\begin{equation}
\frac{P^\dagger [X^\dagger_N , Y_N]}{P[X_N, Y_N]} = \exp \left( - \sigma [X_N, \Lambda_N (Y_N)] - I_c [X_N: Y_N] \right),
\label{fluctuation_f2}
\end{equation}
where the effect of feedback control is involved by the term of  the mutual information that is obtained in the forward experiments.
We stress that, to obtain Eq.~(\ref{fluctuation_f2}),  we do not perform feedback control in the backward experiments.  We just reverse forward protocol as Eq.~(\ref{backward_control}) in the backward experiments.
The same result for a special case was obtained in Ref.~\cite{Horowitz}.
The investigation of the detailed fluctuation theorem in the situations in which feedback control is also performed in the backward processes~\cite{Ponmurugan} is an interesting future challenge.  Such situations would be relevant to, for example, autonomous systems consisting of the controlled system and the controller, in which feedback control should also be needed for the backward processes.  We can expect that the backward processes with feedback control can be used to characterize the reversibility of the autonomous systems.

From the generalized detailed fluctuation theorem (\ref{fluctuation_f2}), we obtain a generalized integral fluctuation theorem~\cite{Sagawa-Ueda3}:
\begin{equation}
\langle e^{-\sigma - I_{\rm c}} \rangle = 1. 
\label{integral_f1}
\end{equation}

Due to the concavity of the exponential function, we obtain a generalized second law of thermodynamics~\cite{Sagawa-Ueda1,Sagawa-Ueda3}:
\begin{equation}
\langle \sigma \rangle \geq - \langle I_{\rm c} \rangle,
\label{second_f1}
\end{equation}
which means that the entropy production can be negative due to the effect of feedback control (or due to the action of Maxwell's demon), and that the lower bound of the entropy production is bounded by the mutual information $\langle I_{\rm c} \rangle$.   

The reason why the entropy production can be negative is that one can rectify the thermal fluctuations by feedback control.  This negative entropy production is compensated for by the excess entropy production in the demon or the feedback controller~\cite{Sagawa-Ueda2}, and therefore the entropy production in the total system consisting of the demon and the information heat engine is consistent with the second law of thermodynamics.
The key feature of feedback control is that it enables us to control the entropy production of a partial system by utilizing the mutual information beyond the limitation of the conventional thermodynamics.
 Inequality~(\ref{second_f1}) identifies the lower bound of the entropy production with feedback control, which plays a role parallel to the conventional second law of thermodynamics that gives the lower bound of zero in the absence of feedback control.  Therefore, inequality~(\ref{second_f1}) is regarded as a generalization of the second law of thermodynamics that can be applied to feedback-controlled processes.

We also obtain, by taking the ensemble average of the logarithm of the both sides of Eq.~(\ref{fluctuation_f2}), that
\begin{equation}
\langle \sigma \rangle + \langle I_{\rm c} \rangle  = 
\int dX_N dY_N P[X_N, Y_N] \ln \frac{P[X_N, Y_N]}{P^\dagger [X^\dagger_N , Y_N]},
\label{relative_f1}
\end{equation}
which is a generalization of the Kawai-Parrondo-Broeck (KPB) equality~(\ref{relative1}).
We note that the right-hand side of  Eq.~(\ref{relative_f1}) is positive because it is the Kullback-Leibler divergence between two probability distributions $P[X_N, Y_N]$ and $P^\dagger [X^\dagger_N , Y_N]$;  thus, inequality (\ref{second_f1}) is reproduced.   We note that equality in (\ref{second_f1}) is achieved if and only if $\sigma + I_{\rm c}$ does not fluctuate, or equivalently,  if
\begin{equation}
P[X_N, Y_N] = P^\dagger [X^\dagger_N , Y_N]
\end{equation}
holds, which implies the reversibility with feedback control~\cite{Horowitz2}.  The more the probability distribution of the forward processes with feedback is different from that of the backward processes without feedback, the more $\langle \sigma \rangle$ is different from $- \langle I_{\rm c} \rangle$.

If the joint distribution of $\sigma$ and $I_{\rm c}$ is Gaussian, we have a generalized fluctuation-dissipation theorem from the second cumulant of Eq.~(\ref{integral_f1}):
\begin{equation}
\langle \sigma + I_{\rm c} \rangle = \frac{1}{2} ( \langle (\sigma + I_{\rm c})^2 \rangle - \langle \sigma + I_{\rm c} \rangle^2 ),
\label{FDT_f1}
\end{equation}
which suggests that there is a trade-off relation between the entropy production and the mutual information.

For the case in which $\sigma = \beta (W - \Delta F)$, Eq.~(\ref{fluctuation_f2}) leads to a generalized Jarzynski equality:
\begin{equation}
\langle e^{- \beta (W - \Delta F) - I_{\rm c}} \rangle = 1, 
\label{Jarzynski_f1}
\end{equation}
and inequality (\ref{second_f1}) leads to
\begin{equation}
\langle W \rangle \geq  \Delta F - k_{\rm B}T\langle I_{\rm c} \rangle.
\label{second_f2}
\end{equation}
We note that Eq.~(\ref{Jarzynski_f1}) and inequality (\ref{second_f2}) are the generalizations of the results obtained in Refs.~\cite{Sagawa-Ueda3,Horowitz}.
By defining $W_{\rm ext} := -  W$  and setting $\Delta F = 0$, we may rewrite inequality (\ref{second_f2}) as 
\begin{equation}
\langle W_{\rm ext } \rangle \leq  k_{\rm B}T\langle I_{\rm c} \rangle,
\label{second_f3}
\end{equation}
which implies that we can extract a positive work up to the term that is equal to the mutual information multiplied by $k_{\rm B}T$, from a thermodynamic cycle with a single heat bath with the assistance of feedback control or Maxwell's demon.  The mutual information can be used as a ``resource'' of the work or the free energy.
In the case of the Szilard engine, $\langle I_{\rm c} \rangle = \langle I \rangle = \ln 2$ and $\langle W_{\rm ext } \rangle = k_{\rm B}T \ln 2$ hold, and therefore the equality in   (\ref{second_f3})  is achieved.  In fact, in the Szilard engine, $\sigma + I = \beta (W - \Delta F ) + I$ does not fluctuate, but is zero for both outcomes ``left'' and ``right.''

We note that, to obtain Eq.~(\ref{integral_f1}) or (\ref{Jarzynski_f1}) experimentally or  numerically,  the condition of $P_{\rm c}[Y_N | X_N] \neq 0$ needs to be satisfied for all $(X_N, Y_N)$.  To explicitly see this, we write $P_{\rm c}[Y_N | X_N] =: \varepsilon >0$.  We then obtain
\begin{equation}
P[X_N, Y_N] e^{-\sigma - I_{\rm c}} = \varepsilon P[X_N] \cdot \frac{1}{\varepsilon} e^{-\sigma + \ln P[Y_N]}, 
\end{equation}
which does not converge to zero with the limit of $\varepsilon \to 0$. On the other hand, in real experiments or numerical simulations, the events with $P[X_N, Y_N]  = 0$ never occur.  Therefore, if $P_{\rm c}[Y_N | X_N] =0$ holds for some $(X_N, Y_N)$,  the  terms associated with zero-probability events make non-zero contributions to  Eq.~(\ref{integral_f1}) and (\ref{Jarzynski_f1}); in such cases, we cannot obtain Eq.~(\ref{integral_f1}) or (\ref{Jarzynski_f1}) experimentally or numerically.  
On the contrary, 
\begin{equation}
P[X_N, Y_N] I_{\rm c}^n= \varepsilon P[X_N] \cdot \left( \ln \frac{\varepsilon}{P[Y_N]} \right)^n
\end{equation}
converges to zero for all $n = 1, 2, \cdots$, in the limit of $\varepsilon \to 0$.  Therefore, we can find $\langle I_{\rm c}^n \rangle$  experimentally and numerically even if $P_{\rm c}[Y_N | X_N] =0$  for some $(X_N, Y_N)$, and also obtain Eqs.~(\ref{relative_f1}), (\ref{FDT_f1}), and inequalities (\ref{second_f1}), (\ref{second_f2}),  (\ref{second_f3}).

\subsection{Generalized Fluctuation Theorem with Efficacy Parameter}

We next derive a different type of nonequilibrium equality. 
In this subsection, we assume that the measurements are Markovian (i.e., $P[y_n | X_n] = P[y_n | x_n]$ holds). 
We perform forward experiments with measurements at times $t_{n_1}, t_{n_2}, \cdots, t_{n_M}$ with feedback control, and perform backward experiments without feedback but only with measurements  at times  $t_{N-n_M}, t_{N-n_{M-1}}, \cdots, t_{N-n_1}$.  

 Let $Y'_N := (y'_{N-n_M}, y'_{N-n_{M-1}}, \cdots, y'_{N- n_1})$ be the measurement outcomes in the backward measurements.  Then, the probability of obtaining $Y'_N$ under the condition of $X_N^\dagger$ is given by 
\begin{equation}
P_{\rm c} [Y'_N | X_N^\dagger] := \prod_{k=1}^M P[y'_{N-n_k} | x_{N-n_k}^\dagger].
\end{equation}
Therefore, the probability of obtaining $Y'_N$ under protocol $\Lambda (Y_N)^\dagger$ is
\begin{equation}
P [Y'_N | \Lambda_N (Y_{N-1})^\dagger] = \int P_{\rm c} [Y'_N | X_N^\dagger] P[X_N^\dagger | \Lambda_N (Y_{N-1})^\dagger] dX_N^\dagger,
\end{equation}
which is normalized as
\begin{equation}
\int P [Y'_N | \Lambda_N (Y_{N-1})^\dagger]  dY'_N = 1,
\label{n1}
\end{equation}
where the probability variable $Y'_N$ is independent of $Y_N$. 

We then consider the time-reversed sequence of $Y_N$.  Let $y_n^\ast$ be the time-reversal of $y_n$; for example, if we measure the momentum, then $y_n^\ast = - y_n$.  We write $Y_N^\dagger  := (y^\ast_{N-n_M}, y^\ast_{N-n_{M-1}}, \cdots, y^\ast_{N- n_1})$. The probability of $Y'_N = Y_N^\dagger$ in the backward experiments is given by
\begin{equation}
P[Y_N^\dagger | \Lambda_N (Y_{N-1})^\dagger] = \int P_{\rm c} [Y_N^\dagger | X_N^\dagger] P[X_N^\dagger | \Lambda_N (Y_{N-1})^\dagger]dX_N^\dagger,
\label{prob2}
\end{equation}
which is the probability of obtaining the time-reversed outcomes by time-reversed measurements during the time-reversed protocol.  We stress that 
\begin{equation}
\int P[Y_N^\dagger | \Lambda_N (Y_{N-1})^\dagger] dY_N^\dagger \neq 1
\end{equation}
in general because  $Y_N^\dagger$ is no longer independent of $Y_{N-1}$.

In the following, we assume that the measurements have the time-reversed symmetry
\begin{equation}
P[y_n^\ast | x_n^\ast] = P[y_n | x_n].
\end{equation}
for all $n$, which leads to
\begin{equation}
P_{\rm c}[Y_n^\dagger | X_n^\dagger] = P_{\rm c}[Y_n | X_n]. 
\label{time_symmetry}
\end{equation}
We then have the ``renormalized'' (or  ``coarse-grained'') detailed fluctuation theorem~\cite{Kawai,Sagawa-Ueda3}
\begin{equation}
\frac{P[Y_N^\dagger | \Lambda (Y_N)^\dagger]}{P[Y_N]} = e^{- \sigma' [Y_N]}, 
\label{fluctuation_f3}
\end{equation}
where $\sigma' [Y_N]$ is  the ``renormalized'' (or ``coarse-grained'') entropy production defined as
\begin{equation}
\begin{split}
\sigma' [Y_N] &:= -\ln \langle e^{-\sigma} \rangle_{Y_N} \\
&= -\ln \int dX_Ne^{-\sigma [X_N, \Lambda_N(Y_{N-1})] }P[X_N | Y_N].
\end{split}
\label{entropy_f2}
\end{equation}
Equality (\ref{fluctuation_f3}) implies that the detailed fluctuation theorem retains its form under the coarse-graining, if we introduce the appropriate coarse-grained entropy production.  
From the concavity of the exponential function, we obtain $\sigma' [Y_N]  \leq \langle \sigma \rangle_{Y_N}$ and $\langle \sigma' \rangle \leq \langle \sigma \rangle$.  The same result for a different setup has been obtained in~\cite{Kawai,Marin}.

The proof of (\ref{fluctuation_f3}) goes as follows.   From the definition of $\sigma' [Y_N]$ and the detailed fluctuation theorem (\ref{fluctuation_f1}), we have
\begin{equation}
\begin{split}
e^{-\sigma' [Y_N]} &= \int dX_N \frac{P[X_N^\dagger | \Lambda_N (Y_{N-1})^\dagger]}{P[X_N | \Lambda_N (Y_{N-1})]} P[X_N | Y_N] \\
&= \int dX_N \frac{P[X_N^\dagger | \Lambda_N (Y_{N-1})^\dagger]}{P[X_N | \Lambda_N (Y_{N-1})]} \frac{P[X_N, Y_N]}{P[Y_N]} \\
&= \frac{1}{P[Y_N]} \int dX_N  P[X_N^\dagger | \Lambda_N (Y_{N-1})^\dagger] P_{\rm c}[Y_N | X_N] \\
&= \frac{1}{P[Y_N]} \int dX_N  P[X_N^\dagger | \Lambda_N (Y_{N-1})^\dagger] P_{\rm c}[Y_N^\dagger | X_N^\dagger]. 
\end{split}
\end{equation}
In the last line, we used the time-reversal symmetry (\ref{time_symmetry}) of the measurements.
By noting Eq.~(\ref{prob2}), we obtain (\ref{fluctuation_f3}).

We note that Eq.~(\ref{fluctuation_f3}) holds regardless of the presence of feedback control.  Without feedback control, Eq.~(\ref{fluctuation_f3}) reduces to
\begin{equation}
\frac{P^\dagger [Y_N^\dagger]}{P[Y_N]} = e^{- \sigma' [Y_N] }.
\label{fluctuation_f4}
\end{equation}

By taking the ensemble average of both sides of Eq.~(\ref{fluctuation_f3}) and noting that $\langle e^{-\sigma'} \rangle = \langle e^{-\sigma} \rangle$ holds, we obtain the second generalization of the integral fluctuation theorem~\cite{Sagawa-Ueda3}
\begin{equation}
\langle e^{-\sigma} \rangle = \gamma,
\label{integral_f2}
\end{equation}
where $\gamma$ is the efficacy parameter of feedback control defined as
\begin{equation}
\gamma := \int P [Y_N^\dagger | \Lambda_N (Y_{N-1})^\dagger] dY_N^\dagger,
\label{gamma}
\end{equation}
which is the sum of probabilities of obtaining the time-reversed outcomes by the time-reversed measurements during the  time-reversed protocols (see FIG.~3).   If $\sigma = \beta (W - \Delta F)$ holds, Eq.~(\ref{integral_f2}) leads to the second generalization of the Jarzynski equality~\cite{Sagawa-Ueda3}:
\begin{equation}
\langle e^{-\beta (W - \Delta F)} \rangle = \gamma.
\label{Jarzynski_f2}
\end{equation}

If the feedback control in the forward processes is ``perfect,'' the particle is expected to return to its initial state with unit probability in the backward processes.
In such a case, $\gamma$ takes the maximum value that equals the number of possible outcomes of $Y_N$.
In fact, for the case of the Szilard engine, $\gamma = 2$ holds corresponding to $W = - k_{\rm B} T \ln 2$ and $\Delta F = 0$~\cite{Sagawa-Ueda3}.  
In contrast, without feedback control, $\gamma$ reduces to $1$ as
\begin{equation}
\gamma := \int P[Y_N^\dagger] dY_N^\dagger = 1,
\end{equation}
which vindicates  the original integral fluctuation theorem. 
Therefore, the measurements in the backward processes are used to characterized to the efficacy of feedback control in the forward processes.

\begin{figure}[htbp]
 \begin{center}
 \includegraphics[width=80mm]{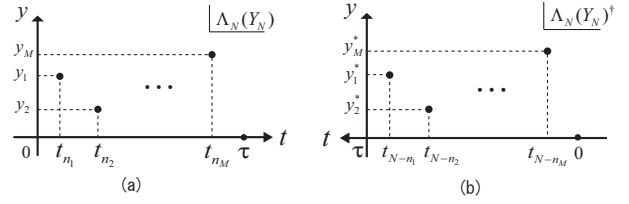}
 \end{center}
 \caption{(a) Forward outcomes $Y_N$ with forward protocol $\Lambda_N (Y_N)$.  (b) Backward outcomes $Y_N^\dagger$ with backward protocol ($\Lambda_N(Y_N)^\dagger$).} 
\end{figure}

We stress that $\sigma$ and $\gamma$ can be measured independently, because $\sigma$ is obtained from the forward experiments with feedback and $\gamma$ is obtained from the backward experiments without feedback.  Therefore, Eqs.~(\ref{integral_f2}) and (\ref{Jarzynski_f2}) can be directly verified in experiments.  In fact, Eq.~(\ref{Jarzynski_f2}) has been verified in a real experiment by using a feedback-controlled ratchet with a Brownian particle~\cite{Toyabe3}.

From Eq.~(\ref{fluctuation_f2}), we have the second generalization of the second law of thermodynamics 
\begin{equation}
\langle \sigma \rangle \geq - \ln \gamma .
\label{second_f4}
\end{equation}
The equality in inequality (\ref{second_f4}) is achieved if $\sigma$ does not fluctuate.
We note that, if the distribution of $\sigma$ is Gaussian, we have a generalized fluctuation-dissipation theorem
\begin{equation}
\langle \sigma \rangle + \ln \gamma = \frac{1}{2} (\langle \sigma^2 \rangle - \langle \sigma \rangle^2). 
\label{FDT_f2}
\end{equation}

While the first generalization (\ref{integral_f1}) only involves the term of the obtained information, the second generalization (\ref{integral_f2}) involves the term of feedback efficacy.  To understand the relationship between the mutual information $I_{\rm c}$ and the feedback efficacy $\gamma$, we introduce the notation
\begin{equation}
C[A] := -\ln \langle e^{-A} \rangle
\end{equation}
for any probability variable $A$.  We note that, if $A$ can be written as $A = tA'$ with $t$ being a real number and $A'$ being another probability variable, then $C[A]$ is the cumulant generation function of $A'$.  By using  this notation, we have
\begin{equation}
C[\sigma] + C [I_{\rm c}] - C [\sigma + I_{\rm c}] = - \ln \gamma,
\label{correlation1}
\end{equation}
because $C[\sigma] = -\ln \gamma$  in Eq.~(\ref{integral_f2}), $C[I_{\rm c}] = 0$ holds as in Eq.~(\ref{mutual_integral}), and $C[\sigma +I_{\rm c}] = 0$ holds as in Eq.~(\ref{integral_f1}).  Equality (\ref{correlation1}) implies that $-\ln \gamma$ is a measure of the correlation between $\sigma$ and $I_{\rm c}$.  This can be more clearly seen by the cumulant expansion of Eq.~(\ref{correlation1}) if the joint distribution of $\sigma$ and $I_{\rm c}$ is Gaussian:
\begin{equation}
\langle \sigma I_{\rm c} \rangle - \langle \sigma \rangle \langle I_{\rm c} \rangle = - \ln \gamma.
\label{correlation2} 
\end{equation}  
Therefore, $\gamma$ characterizes how efficiently we use the obtained information to decrease the entropy production by feedback control:  if $\gamma$ is large, the more $I_{\rm c}$ we obtain, the less $\sigma$ is.

We can also derive another  nonequilibrium equality which also gives us the information about the feedback efficacy.   By taking logarithm of the both sides of Eq.~(\ref{fluctuation_f2}), we obtain 
\begin{equation}
\langle \sigma' \rangle = \int dY_N P[Y_N] \ln \frac{P[Y_N]}{P[Y_N^\dagger | \Lambda_N (Y_{N-1})^\dagger]},
\label{relative_f2}
\end{equation}
which is a generalization of Eq.~(\ref{relative1}).  The same result under a different situation has also been obtained in Ref~\cite{Kawai,Marin}.  Equality (\ref{relative_f2}) implies that the renormalized entropy production equals the Kullback-Leibler divergence-like quantity between the forward probability $P[Y_N]$ and the backward probability $P[Y_N^\dagger | \Lambda_N (Y_{N-1})^\dagger]$.  In fact, without feedback control, the right-hand side of Eq.~(\ref{relative_f2}) reduces to the Kullback-Leibler divergence between  $P[Y_N]$ and  $P^\dagger[Y_N^\dagger]$ and therefore the both sides of Eq.~(\ref{relative_f2}) are positive, which is consistent with the second law of thermodynamics.   On the contrary, in the presence of feedback control, the right-hand side is no longer the Kullback-Leibler divergence, because $P[Y_N^\dagger | \Lambda_N (Y_{N-1})^\dagger]$ is not  a normalized probability distribution in terms of $Y_N^\dagger$.  Therefore the both sides of (\ref{relative_f2}) can be negative. 
Since $\langle \sigma' \rangle \leq \langle  \sigma \rangle$, the entropy production $\langle \sigma \rangle$ is bounded from below by the right-hand side of Eq.~(\ref{relative_f2}):
\begin{equation}
\langle \sigma \rangle \geq \int dY_N P[Y_N] \ln \frac{P[Y_N]}{P[Y_N^\dagger | \Lambda_N (Y_{N-1})^\dagger]}.
\label{relative_f3}
\end{equation}
Without feedback control, the right-hand side of (\ref{relative_f3}) gives a positive bound, while, with feedback control, the right-hand side can give a negative bound.
We note that, for a quantum generalization of the Szilard engine with multi-particles,  essentially the same result as Eq.~(\ref{relative_f2}) has been obtained~\cite{SWKim}.

We  note that special cases of our results in this section were obtained  elsewhere.  We have derived  two types of the generalized Jarzynski equality for the cases with a single measurement in the presence of  a single heat bath in Ref.~\cite{Sagawa-Ueda3}.   In Ref.~\cite{Horowitz}, the detailed fluctuation theorem and the Jarzynski equality were obtained for the cases with multi-measurements and feedback in the presence of a single heat bath.  In Ref.~\cite{Suzuki2}, a generalized Jarzynski equality was also obtained for the Kalman filter and the optimal control.  The results in this paper include all of the above results, and generalize them to the cases of multi-heat baths and non-Markovian measurements.    

We also note that the generalized Jarzynski equality~(\ref{Jarzynski_f2}) with a single measurement was experimentally verified by using a feedback-controlled ratchet with a colloidal particle~\cite{Toyabe3}.
Moreover, Eq.~(\ref{Jarzynski_f2}) has been generalized to quantum systems~\cite{Morikuni}.

\section{Examples}

We now discuss two examples which illustrate the essential features of our general results.  We analytically discuss a generalized Szilard engine with measurement errors  in Sec.~VI A, and numerically discuss a feedback-controlled ratchet in Sec.~VI B.

\subsection{Szilard Engine with Measurement Errors}

As an example with a classical measurement, we discuss a generalized Szilard engine with measurement errors, which will be shown to achieve the upper bound of inequality (\ref{second_f2})or (\ref{second_f3}) for an arbitrary error rate. The control protocol of the generalized Szilard engine is given by the following steps, which are described FIG.~4. 

\textit{Step 1: Initial state.}
 A single-particle classical gas is in a box.  The initial state of the gas is in thermal equilibrium  with a single heat bath at temperature $T = (k_{\rm B}\beta)^{-1}$.

\textit{Step 2: Insertion of the barrier.} We insert a barrier in the middle of the box, and divide it to two boxes with the same volume.  Here, we do not know in which box the particle is.  For simplicity of notations, we write   ``left'' as ``$0$'' and ``right'' as ``$1$.''  In other words, the position $x$ of the particle  is given by $x= 0$ or $x=1$.    We do not need any work during this process as proved in~\cite{SWKim}.

\textit{Step 3: Measurement.} We measure the position of the particle.  We assume that the measurement is equivalent to the binary symmetric channel with error rate $\varepsilon$~\cite{Cover-Thomas}; the measurement outcome takes  $y=0$ or $1$, and  the measurement error is characterized by conditional probabilities $P[0|0] = P[1|1] - 1 = \varepsilon$ and $P[0|1] = P[1|0] = \varepsilon$ with $0 \leq \varepsilon \leq 1$.   We note that $x=y$ holds for the original Szilard engine without error ($\varepsilon = 0$).  

\textit{Step 4: Feedback.}
We next move the position of the barrier quasi-statically and isothermally.  The protocol of moving the barrier depends on measurement outcome $y$.  
Let $v_0$ ($0 \leq v_0 \leq 1$) and $v_1$ ($0 \leq v_0 \leq 1$) be real numbers.
We assume that, after we move the barrier, the ratio of the volumes of the boxes is assumed to be $v_0 : 1-v_0$ for $y=0$, or $1-v_1 : v_1$ for $y=1$.  We note that, in the case of the original Szilard engine, $v_0 = v_1 = 1$ holds.  In this process, we extract the work from the engine.  The amounts of the work are given by $k_{\rm B} T \ln 2 v_0$ if $(x, y) = (0,0)$, $k_{\rm B}  T\ln 2 (1- v_0)$ if $(x, y) = (0,1)$, $k_{\rm B }T \ln 2 (1-v_1)$ if $(x, y) = (1,0)$, and $k_{\rm B}  T\ln 2 v_1$ if $(x, y) = (1,1)$.  The feedback protocol is characterized by  $v_0$ and $v_1$.

\textit{Step 5: Removal of the barrier.}
We remove the barrier without any work.  The engine then returns to the initial state.

\begin{figure}[htbp]
 \begin{center}
  \includegraphics[width=80mm]{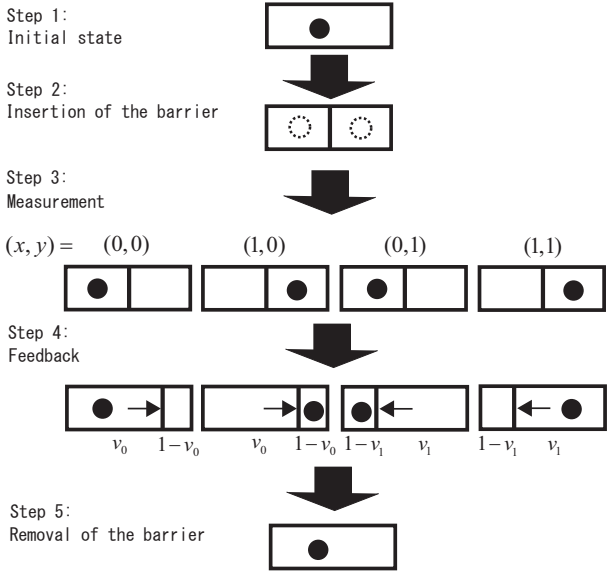}
 \end{center}
 \caption{Generalized Szilard engine with measurement error rate $\varepsilon$, where  $x$ denotes the states of the system, and $y$ denote the measurement outcomes.  The control protocol is determined by $y$.} 
\end{figure}

From the total process, we extract the average work
\begin{equation}
\begin{split}
\langle W_{\rm ext} \rangle = &k_{\rm B}T \Bigl(  \ln 2 + \frac{1- \varepsilon}{2} \ln v_0 + \frac{\varepsilon}{2} \ln (1- v_0)   \\
&+ \frac{\varepsilon}{2} \ln ( 1- v_1)  + \frac{1 - \varepsilon}{2} \ln v_1 \Bigr).
\end{split}
\label{Szilard_work}
\end{equation}
We note that $\Delta F^{\rm S} = 0$ holds.  
We then maximize $\langle W_{\rm ext} \rangle$ under a given measurement error $\varepsilon$ by changing  $v_0$ and $v_1$.
The maximum value of $\langle W_{\rm ext} \rangle$ is achieved when 
\begin{equation}
v_0 = v_1 = 1 - \varepsilon,
\end{equation}
for which the maximum work is given by 
\begin{equation}
\langle W_{\rm ext} \rangle  =  k_{\rm B}T [ \ln 2 + \varepsilon \ln \varepsilon + (1 - \varepsilon) \ln (1 - \varepsilon)].
\label{Szilard_work_opt}
\end{equation}
On the other hand, the mutual information of the binary symmetric channel is given by
\begin{equation}
\langle I \rangle  = \ln 2 + \varepsilon \ln \varepsilon + (1 - \varepsilon) \ln (1 - \varepsilon).
\end{equation}
Therefore, we obtain
\begin{equation}
\langle W_{\rm ext} \rangle = k_{\rm B}T  \langle I \rangle ,
\end{equation}
which means that the generalized Szilard engine achieves the upper bound of the extractable work (\ref{second_f2}) or (\ref{second_f3}) for any amount of the mutual information.


We also check the generalized Jarzynski equalities in this model for arbitrary $v_0$, $v_1$, and $\varepsilon$.
We first note that $I [x:y]$ is given by $\ln 2 (1 - \varepsilon)$ when $(x, y) = (0,0)$, $\ln 2 \varepsilon$ when $(x, y) = (0,1)$, $\ln 2 \varepsilon$ when $(x, y) = (1,0)$, and $\ln 2 (1-\varepsilon)$ when $(x, y) = (1,1)$.   Therefore we obtain
\begin{equation}
\langle e^{-\beta W - I} \rangle = \frac{v_0 + (1-v_0) + (1-v_1) + v_1}{2} = 1,
\end{equation}
which confirms Eq.~(\ref{Jarzynski_f1}).

We next consider the second generalization (\ref{Jarzynski_f2}) of the Jarzynski equality.  Corresponding to two measurement outcomes $y= 0,1$, we have two backward control protocols as follows (see also FIG.~5).

\textit{Step 1. Initial state.}  The initial state of the backward control is in the thermal equilibrium.

\textit{Step 2. Insertion of the barrier.}  Corresponding to Step 5 of the forward process, we insert the barrier and decide the box into two boxes, because the time-reversal of the barrier removal is the barrier insertion.  Corresponding to $y=0$ or $y=1$ in the forward process, we divide the box with the ratio $v_0 : 1 - v_0$ or $1-v_1: v_1$, respectively.

\textit{Step 3. Moving the barrier.}  We next move the barrier to the middle of the box quasi-statically and isothermally.  This is the time-reversal of the feedback control in Step 4 of the forward process.

\textit{Step 4.  Measurement.}  We perform the measurement to find in which box the particle is in.  Corresponding to the backward protocol with $y=0$, we obtain the outcomes of backward measurement $y' = 0$ with probability $P [ y'= 0 | \Lambda (y=0)^\dagger] = v_0 (1 - \varepsilon) + (1 - v_0) \varepsilon$ and $y' = 1$ with probability $P [ y'= 1 | \Lambda (y=0)^\dagger] = v_0  \varepsilon + (1 - v_0) ( 1-  \varepsilon)$.  On the other hand, corresponding to the backward protocol with $y=1$, we obtain the outcomes of backward measurement $y' = 0$ with probability $P [ y'= 0 | \Lambda (y=1)^\dagger] = v_1  \varepsilon + (1 - v_1) (1- \varepsilon)$ and $y' = 1$ with probability $P [ y'= 1 | \Lambda (y=1)^\dagger] = v_1 ( 1-  \varepsilon ) + (1 - v_0)   \varepsilon$.

\textit{Step 5. Removal of the barrier.}
We remove the barrier  and the system returns to the initial state.  This is the time-reversal of the barrier insertion in Step 2 of the forward process.

\begin{figure}[htbp]
 \begin{center}
  \includegraphics[width=75mm]{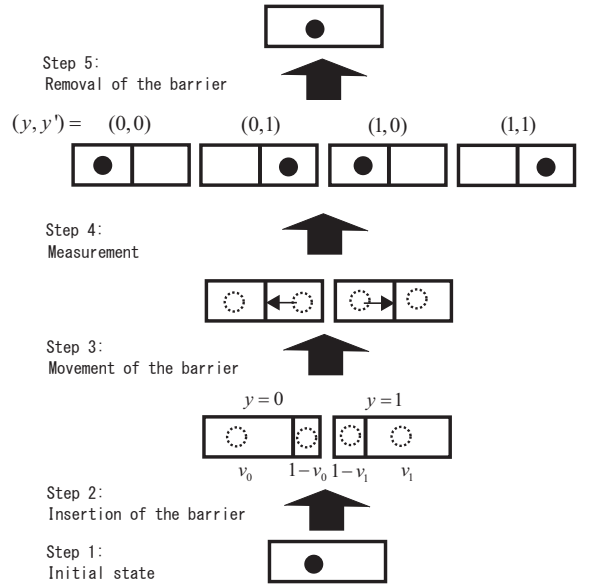}
 \end{center}
\caption{Backward processes of the generalized Szilard engine.  Corresponding to $y$ that denotes the measurement outcomes in the forward process, we have two control protocols in the backward process, where  $y'$ denotes the measurement outcomes in the backward process.} 
\end{figure}

From Step 4 of the backward process, we have 
\begin{equation}
\begin{split}
\gamma &:= P [ y'= 0 | \Lambda (y=0)^\dagger]  + P [ y'= 1 | \Lambda (y=1)^\dagger]  \\
&= (1 - \varepsilon) (v_0 + v_1) + \varepsilon (2 - v_0 - v_1).
\end{split}
\end{equation}
On the other hand, we can straightforwardly obtain 
\begin{equation}
\langle e^{-\beta W} \rangle = (1 - \varepsilon) (v_0 + v_1) + \varepsilon (2 - v_0 - v_1),
\end{equation}
which confirms Eq.~(\ref{Jarzynski_f2}).


\subsection{Feedback-controlled Ratchet}

We next discuss a model for Brownian motors~\cite{Vale,Prost,Parrondo2,Reimann,Hanggi1,Hanggi2}, in particular a feedback-controlled ratchet~\cite{Cao1,Lopez,Cao3}.
We consider a rotating Brownian particle with a periodic boundary condition.  
Let $x$ be the position or the angle of the particle, and its boundary condition is given by $x = x+L$ with $L$ being a constant.  
 In the following, we restrict the particle's position to $-L/2 \leq x < L/2$. 
We assume that the particle obeys the overdamped Langevin equation Eq.~(\ref{Langevin}), and that control parameter $\lambda$ takes two values ($\lambda= 0$ or $1$).  Corresponding to them,  the ratchet potential $V$ takes the following two profiles (FIG.~6):
\begin{eqnarray}
V(x, 0)  &= &
\left\{
\begin{array}{l}
K (x+ L/2) / l \\
\ \ \ \ \ \ \  (-L/2 \leq x < -L/2 + l), \\
- K (x - L/2) / (L-l) \\
\ \ \ \ \  \ \ (-L/2 + l \leq x < L),
\end{array}
\right. \\
V(x, 1)  &=&
\left\{
\begin{array}{l}
- K (x + L/2 - l) / (L-l) \\
\ \ \ \ \  \ \ (-L/2 \leq x < -L/2 + l), \\
K (x+L/2 + 2l) / l \\
\ \ \ \ \  \ \  (-L/2 + l \leq x < -L/2 + 2l), \\
- K (x - L/2 - l) / (L-l) \\
\ \ \ \ \ \ \  (-L/2 + 2l \leq x < L/2),
\end{array}
\right. 
\end{eqnarray}
where $l$ is a constant with $0 < l < L/2$, and $K$ is a positive constant that characterizes the height of the potential.
\begin{figure}[htbp]
 \begin{center}
  \includegraphics[width=80mm]{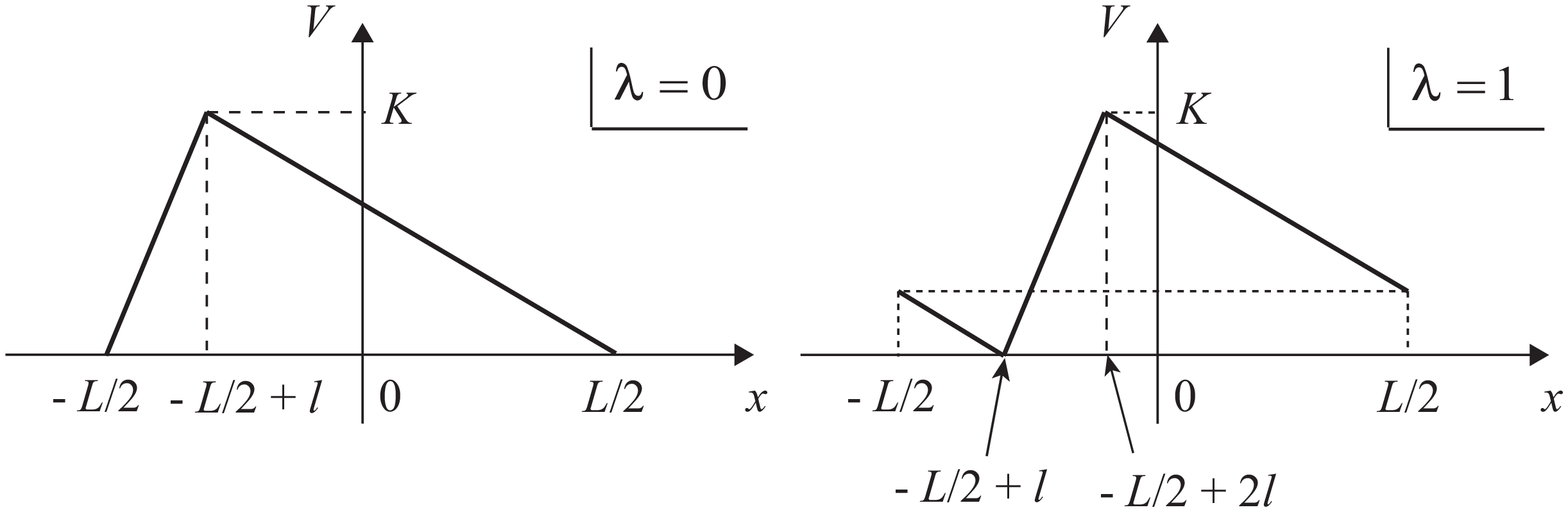}
 \end{center}
 \caption{Two shapes of potential $V(x, \lambda)$ corresponding to $\lambda = 0, 1$.} 
\end{figure}

We start with the initial equilibrium with parameter $\lambda = 0$, and control the system  from time $t = 0$ to $\tau$ with the following three  protocols.
\begin{enumerate}
\item \textit{Trivial control.} We do not change the parameter $\lambda = 0$.
\item \textit{Flashing ratchet.} At times $t = m \tau_0$ with $m$ being integers and $\tau_0$ being a constant, we switch parameter  $\lambda$ from $0$ to $1$ or from $1$ to $0$ periodically. 
\item \textit{Feedback-controlled ratchet.} At times $t = m \tau_0$, we switch the parameter with the following feedback protocol.  We measure the position $x$ at  $t = m \tau_0$  without error. We then  set $\lambda = 1$ from $t = m \tau_0$ to $(m+1) \tau_0$ if and only if the outcome is in $-L/2 \leq x < -L/2 + l$.  Otherwise, parameter $\lambda$ is set to $0$. 
\end{enumerate}

For numerical simulations, we set  $l = 3L/10$, $K = 3k_{\rm B}T$, $\tau_0 = 0.05$, and $\tau = 0.25$, with units $k_{\rm B}T = 1$, $L = 1$, and $\eta / 2 = 1$.  We performed the simulations by discretizing Eq.~(\ref{Langevin}) with $\Delta t = 0.00025$ for $1,000,000$ samples.   We note that, to obtain the initial thermal equilibrium, we waited $\tau_{\rm wait} = 0.5$ and checked that the system was fully thermalized in the periodic ratchet with parameter $\lambda = 0$. 

 The time evolution of the ensemble average $\langle x (t) \rangle$ is plotted in FIG.~7 (a) for the above three protocols.  As expected, nothing happens for the first protocol, while the particle is transported  to  the right on average for the second  and third protocols.  In the case of the feedback-controlled ratchet, the particle is transported to the right faster than the case of the flashing ratchet.  Figure 7 (b)  shows the time evolution of the work $\langle W(t) \rangle$ that is performed on the particle.  The work is induced only in the switching times.  We find that, in order to transport the particle, the energy input to the particle with feedback control is smaller than that with the flashing. 
 
\begin{figure}[htbp]
\begin{center}
\includegraphics[width=80mm]{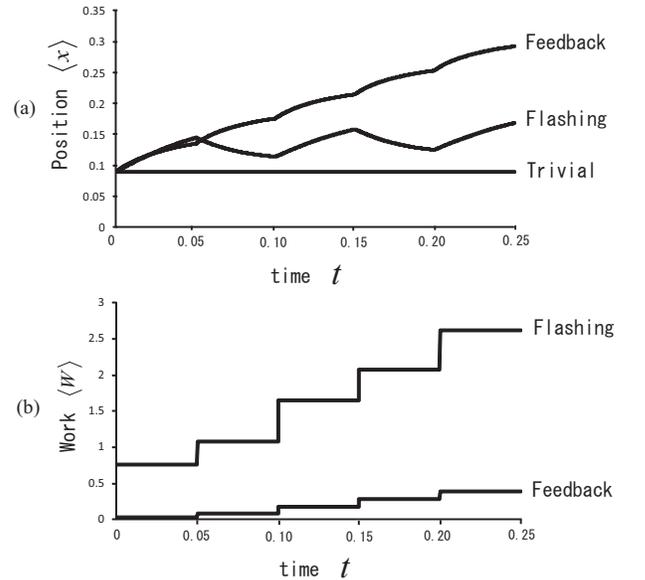}
\end{center}
\caption{(a) Numerical results of the ensemble average of trajectory $x(t)$ corresponding to the three control protocols: the trivial control, the flashing ratchet, and the feedback-controlled ratchet.  (b) Numerical result of the ensemble average  of the work $W(t)$ corresponding to the flashing ratchet and the feedback-controlled ratchet.} 
\end{figure}

Figure 8 shows the left-hand side of the Jarzynski equality $\langle e^{-\beta (W - \Delta F)} \rangle$ for the flashing and feedback-controlled ratchet, and the efficacy parameter $\gamma$ for the feedback-controlled ratchet.  We note that $\Delta F = 0$ always holds.
With feedback control, $\langle e^{-\beta (W - \Delta F)} \rangle$ increases from $1$ as the number of switchings increases,  while, without feedback control, $\langle e^{-\beta (W - \Delta F)} \rangle$ converges to $1$ for all switching times in consistent with the original Jarzynski equality.   On the other hand,  to obtain $\gamma$, we numerically performed the backward experiments.  The discretization of the time is $\Delta t = 0.0005$, and the number of the samples is $10,000$ for each trajectory of  $\lambda (t)$.   We note that the number of the trajectories of $\lambda$ is given by $2^m$ with $m$ times of switchings.  Figure 8 shows a good coincidence between $\langle e^{-\beta W} \rangle$ and $\gamma$, which confirms the validity of Eq.~(\ref{Jarzynski_f2}) in the feedback-controlled ratchet.

\begin{figure}[htbp]
\begin{center}
\includegraphics[width=88mm]{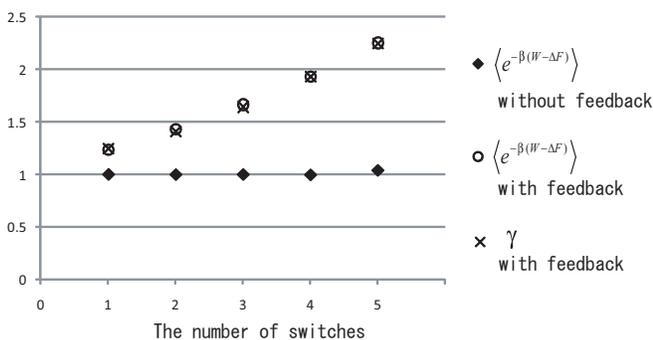}
\end{center}
\caption{Numerical tests of the Jarzynski equality for the flashing ratchet and a generalized Jarzynski equality (\ref{Jarzynski_f2}) for the feedback-controlled ratchet.} 
\end{figure}

\section{Conclusions}

In this paper, we have studied the effects of measurements and feedback control on nonequilibrium thermodynamic systems.
In particular, we have generalized nonequilibrium equalities to the systems that are subject to feedback control.   
Our formulations and results are applicable to a broad class of classical nonequilibrium systems.

In Sec.~II, we reviewed stochastic thermodynamics, by focusing on the nonequilibrium equalities.
In Sec.~III,  we formulated measurements on nonequilibrium systems, and defined  mutual information $I_{\rm c}$ by (\ref{mutual1}) for multi-measurements.   In Sec.~IV, we formulated feedback control on nonequilibrium systems.  We discussed the properties of the joint probability (\ref{joint_prob}), which is well-defined due to causality.
We introduced the mutual information $I_{\rm c}$ by (\ref{mutual_f}), which is not equivalent to $I$ in the presence of feedback control.  In fact, $I_{\rm c}$ describes the correlation between the system and the outcomes, which characterizes the effective information obtained by the measurements.  We have also shown that the detailed fluctuation theorem (\ref{fluctuation_f1}) holds in the presence of feedback control. 

Section~V constitutes the main results of this  paper.  We derived two types of generalizations of the nonequilibrium equalities.  In Sec.~V A, we derived a generalized detailed fluctuation theorem (\ref{fluctuation_f2}) which involves the mutual information. 
Based on Eq.~(\ref{fluctuation_f2}), we derived the generalizations of the integral fluctuation theorem (\ref{integral_f1}), the Jarzynski equality (\ref{Jarzynski_f1}), the second laws (\ref{second_f1}) (\ref{second_f2}) (\ref{second_f3}), the fluctuation-dissipation theorem (\ref{FDT_f1}), and the KPB equality (\ref{relative_f1})  that all involve the mutual information.  In Sec.~V B, we derived the renormalized detailed fluctuation theorem (\ref{fluctuation_f3}), and derived the generalizations of  the integral fluctuation theorem (\ref{integral_f2}), the Jarzynski equality (\ref{Jarzynski_f2}), the second law (\ref{second_f4}), the  fluctuation-dissipation theorem (\ref{FDT_f2}), and the KPB equality (\ref{relative_f2}).  We have shown that mutual information $I_{\rm c}$, rather than $I$, plays the crucial role to formulate the nonequilibrium equalities under feedback control.  These results are the generalizations of the fundamental equalities in nonequilibrium statistical mechanics to feedback-controlled processes, and lead to the generalized second law of thermodynamics with feedback control, which gives the minimal energy cost that is needed for the feedback control.

In Sec.~VI, we discussed simple examples to explicitly show that our results in Sec.~V can be applied to typical situations. In Sec.~VI A, we discussed the Szilard engine with measurement errors that achieves the equality of the generalized second law of thermodynamics (\ref{second_f2}) or (\ref{second_f3}).  This is an important model to quantitatively illustrate that the mutual information can be converted to the work.   We also confirmed the two generalized Jarzynski equalities (\ref{Jarzynski_f1}) (\ref{Jarzynski_f2}) in the generalized Szilard engine.  In Chap.~VI B, we considered a feedback-controlled ratchet and confirmed a generalized Jarzynski equality (\ref{Jarzynski_f2}).

All of our formulations and results are consistent with the original nonequilibrium equalities and the second law of thermodynamics, and our results serve as the fundamental principle of nonequilibrium thermodynamics of feedback control.  We note that, in our results such as Eq.~(\ref{fluctuation_f2}), the thermodynamic quantities and the information contents are treated on an equal footing.  Therefore, our theory may be regarded as the nonequilibrium version of ``information thermodynamics''~\cite{Sagawa-Ueda1,Sagawa-Ueda2}, which serves as the fundamental theory  of  nonequilibrium information heat engines.

\appendix

\section{Physical Meaning of the Entropy Production}

In this appendix, we discuss the physical meanings of the entropy production $\sigma$ in the following two typical setups to clarify the typical situations to which our results apply.

\textit{Isothermal processes.}
We assume that there is a single heat bath at temperature $T = (k_{\rm B}\beta)^{-1}$, and that the initial distributions of both forward and backward  experiments are  in the canonical distributions.  We stress that we do not assume that the final distributions of both the forward and backward experiments are in the canonical distributions: the final distribution of the forward (backward) experiments does not necessarily equal the initial distribution of the backward (forward) experiments.  Let $H(x,\lambda)$ be the Hamiltonian of the system with the time symmetry $H(x, \lambda) = H(x^\ast, \lambda^\ast)$.  The canonical distribution with parameter $\lambda$ is given by
\begin{equation}
P_{\rm can}[x|\lambda] := e^{\beta (F(\lambda)  - H (x, \lambda))},
\end{equation}
where 
\begin{equation}
F(\lambda) := - k_{\rm B}T \ln \int dx e^{-\beta H(x, \lambda)}
\end{equation}
is the Helmholtz free energy.
In this situation, the entropy production reduces to
\begin{equation}
\sigma [X_N] = \beta (W[X_N] - \Delta F),
\end{equation}
where
\begin{equation}
W[X_N] := H(x_N, \lambda_{\rm fin}) -  H(x_0, \lambda_{\rm int})  - Q [X_N]
\end{equation}
is the work performed on the system from the external parameter, and $\Delta F := F(\lambda_{\rm fin}) - F(\lambda_{\rm int})$ is the free-energy difference. 
In this case,  Eq.~(\ref{integral1}) leads to the Jarzynski equality (\ref{Jarzynski1}), and the second law  (\ref{second1})  reduces to inequality (\ref{second_work}).

\textit{Transition between arbitrary nonequilibrium states:}  
We assume that there are several heat baths, and that we can control the strength of interaction between the system and the baths through $\lambda$.  In other words, we can attach or detach the system from the baths by controlling $\lambda$; for example, we can attach an adiabatic wall to the system. 
We set an arbitrary initial distribution $P_0 [x_0]$ for the forward experiments.  On the other hand, the initial state of the backward experiments is assumed to be taken as $P_0^\dagger [x_0^\dagger] := P_N [x_N]$, where $P_N[x_N]$ is the final distribution of the forward experiments.  Although this choice of the backward initial state is artificial and is difficult to be experimentally realized except for special cases, this backward initial state is a theoretically useful tool to derive a version of the second law of thermodynamics as follows.  In this case, the entropy production is given by
\begin{equation}
\sigma [X_N] = - \ln P_N [x_N] + \ln P_0 [x_0] - \sum_i \beta_i Q_i [X_N],
\end{equation}
and its ensemble average leads to
\begin{equation}
\langle \sigma \rangle = S_N - S_0 - \sum_i \beta_i \langle Q_i \rangle,
\end{equation}
where
\begin{equation}
S_n := - \int P_n [x_n] \ln P_n [x_n] dx_n
\end{equation}
is the Shannon entropy  at time $t_n$.  By introducing notation $\Delta S := S_N - S_0$, the second law (\ref{second1}) leads to
\begin{equation}
\Delta S \geq \sum_i \beta_i \langle Q_i \rangle.
\label{steady_state2}
\end{equation}

\begin{acknowledgments}
We are grateful to  Y. Fujitani, H. Hayakawa, H. Hasegawa, J. M. Horowitz, S. Ito, K. Kawaguchi,  T. S. Komatsu, N. Nakagawa, K. Saito, M. Sano, S. Sasa, H. Suzuki, H. Tasaki, and S. Toyabe for valuable discussions. 
This work was supported by a Grant-in Aid for Scientific Research on Innovative Areas ``Topological Quantum Phenomena'' (KAKENHI 22103005) from the Ministry of Education, Culture, Sports, Science and Technology (MEXT) of Japan, and by a Global COE program ``Physical Science Frontier'' of MEXT, Japan. TS acknowledges JSPS Research Fellowships for Young Scientists (Grant No.~208038) and the Grant-in-Aid for Research Activity Start-up (Grant No.~11025807).
\end{acknowledgments}

\end{document}